\documentstyle[preprint,amsmath,amssymb,array,mathtools]{elsarticle}

\newtheorem{Theorem}{{Theorem}}

\newdefinition{Definition}[Theorem]{{Definition}}
\newtheorem{Notation}[Theorem]{{Notation}}
\newtheorem{Lemma}[Theorem]{{Lemma}}
\newtheorem{Corollary}[Theorem]{{Corollary}}
\newtheorem{Proposition}[Theorem]{{Proposition}}
\newtheorem{Conjecture}[Theorem]{{Conjecture}}
\newtheorem{Example}[Theorem]{{Example}}

\newenvironment{Proof}{{\sc Proof:}}{\($\qed$\) \par}

\newcommand{\lcm}{{\textit{lcm}}}
\def\symsuc{\mathrm{Dist}_{\{1\}}}
\newcommand{\ignore}[1]{}

\newcommand{\bu}{\bullet}
\newcommand{\Z}{{\mathbb Z}}

\newcommand{\Diff}{\operatorname{Diff}}
\newcommand{\Dist}{\operatorname{Dist}}
\newcommand\Csp{\operatorname{CSP}}
\newcommand\Aut{\operatorname{Aut}}
\newcommand\suc{\operatorname{succ}}

\begin{document}

\begin{frontmatter}
\title{Distance Constraint Satisfaction Problems\tnoteref{tn1}}

\tnotetext[tn1]{An extended abstract of this paper appeared at MFCS 2010 \cite{BodDalMarPin}.}

\author[tud]{Manuel Bodirsky\fnref{fn1}}


\author[upf]{Victor Dalmau\fnref{fn2}}
\author[mdx]{Barnaby Martin\fnref{fn4}}
\author[tud]{Antoine Mottet\fnref{fn1}}
\author[prag]{Michael Pinsker\fnref{fn3}}

\address[tud]{Institut f\"ur Algebra, TU Dresden, Dresden, Germany}
\address[upf]{Universitat Pompeu Fabra, Barcelona, Spain.}
\address[mdx]{Department of Computer Science, Middlesex University, London, UK.}
\address[prag]{Department of Algebra, MFF UK, Sokolovska 83, 186 00 Praha 8, Czech Republic.}

\fntext[fn1]{Manuel Bodirsky and Antoine Mottet have received support from the European Research Council under the European Community's Seventh Framework Programme (FP7/2007-2013 Grant Agreement no. 257039).}

\fntext[fn2]{Victor Dalmau has been supported by the MCINN grant TIN2010-20967-C04-02.}
\fntext[fn3]{Michael Pinsker is grateful for
support through
Erwin-Schr\"{o}dinger-Fellowship J2742-N18 and projects P21209 and P27600 of the
Austrian Science Fund (FWF), as well as through an APART-fellowship of
the Austrian Academy of Sciences.}
\fntext[fn4]{Barnaby Martin is supported by EPSRC grant EP/L005654/1.}

\begin{abstract}
We study the complexity of constraint satisfaction problems
for templates $\Gamma$ over the integers 
where the relations are first-order definable from the successor function.
In the case that $\Gamma$ is locally finite (i.e., the Gaifman graph of $\Gamma$ has finite degree),
we show that
$\Gamma$ is homomorphically equivalent to a structure
with one of two classes of polymorphisms (which we call modular max and modular min)
and the CSP for $\Gamma$ can be solved in polynomial time, or 
$\Gamma$ is homomorphically equivalent to a finite transitive structure, or
 the CSP for $\Gamma$ is NP-complete.
Assuming a widely believed conjecture from finite domain constraint
satisfaction (we require the \emph{tractability conjecture} by Bulatov, Jeavons and Krokhin in the special case of \emph{transitive} finite templates), 
this proves that those CSPs have a complexity dichotomy, that is, are either in P or NP-complete.  
\end{abstract}


\end{frontmatter}

\section{Introduction}
Constraint satisfaction problems appear naturally in many
areas of theoretical computer science, for example in artificial intelligence, optimization, computer algebra, computational biology,
computational linguistics, and type systems for programming
languages. Such problems are typically NP-hard, but sometimes
they are polynomial-time tractable. The question as to which
CSPs are in P and which are hard has stimulated 
a lot of research in the past 15 years. For pointers to the literature, there is a collection of survey articles~\cite{CSPSurveys}.


The \emph{constraint satisfaction problem CSP} for a fixed 
(not necessarily finite) structure
$\Gamma$ with a finite relational signature $\tau$ 
is the computational problem of deciding whether
a given primitive positive sentence is true in $\Gamma$.
A formula is \emph{primitive positive} if it is of the form 
$\exists x_1,\dots,x_n \, (\psi_1 \wedge \dots \wedge \psi_m)$
where each $\psi_i$ is an \emph{atomic formula} over $\Gamma$,
that is, a formula of the form $y_1=y_2$ or $R(y_1,\dots,y_j)$ 
for a relation symbol 
$R$ of a relation from $\Gamma$. The structure $\Gamma$
is also called the \emph{template} of the CSP.

The class of problems that can be formulated as a CSP for a fixed
structure $\Gamma$ is very large. It can be shown that for every
computational problem there is a structure $\Gamma$ such that the
CSP for $\Gamma$ is equivalent to this problem under polynomial-time
Turing reductions~\cite{BodirskyGrohe}. This makes it very unlikely
that we can give good descriptions of all those $\Gamma$ where
the CSP for $\Gamma$ is in P. 
In contrast, the class of CSPs for a \emph{finite} 
structure $\Gamma$ is
quite restricted, and indeed it has been conjectured
that the CSP for $\Gamma$ is either in P or NP-complete in this case~\cite{FederVardi}. So it appears to be natural to study the CSP for classes 
of infinite structures $\Gamma$ that share good properties with
finite structures.

In graph theory and combinatorics, there are two major 
concepts of \emph{finiteness} for infinite structures. The first
is $\omega$-categoricity: a countable structure is $\omega$-categorical if and only if its automorphism group has for all $n$ only finitely
many orbits in its natural action on $n$-tuples~\cite{Oligo,Marker,Hodges}. 
This property
has been exploited to transfer techniques that were known to
analyze the computational complexity of CSPs with finite domains
to infinite domains~\cite{BodirskyNesetrilJLC,tcsps-journal,BodPin-Schaefer-both}; see also the introduction of~\cite{BodChenPinsker}.

The second concept of finiteness is the property of an infinite graph or structure to be \emph{locally finite} (see Section 8 in~\cite{diestel}). 
A graph is called locally finite if every vertex is contained in a finite number of edges; a relational structure is called locally finite if its Gaifman graph (definition given in Section~\ref{sect:prelims}) is locally finite. Many conjectures that are open for general infinite graphs
become true for locally finite graphs, and many results that
are difficult become easy for locally finite graphs.

In this paper, we initiate the study of CSPs with locally finite
templates by studying locally finite templates $\Gamma$
that have a \emph{first-order definition} in $({\mathbb Z}; \suc)$, that is, $\Gamma$ has the 
domain ${\mathbb Z}$ and all relations of $\Gamma$
can be defined by a first-order formula over the 
successor relation on the integers, 
$\suc = \{(x,y) \; | \; y=x+1\}$. 

As an example, consider the directed graph with vertex set
${\mathbb Z}$ which has an edge between $x$ and $y$
if the difference, $y-x$, between $x$ and $y$ is either $1$ or $3$.
This graph is the structure $({\mathbb Z}; \Diff_{\{1,3\}})$
where $\Diff_{\{1,3\}} = \{(x,y) \; | \; y-x \in \{1,3\} \}$,  which has a first-order definition over $({\mathbb Z}; \suc)$ since $\Diff_{\{1,3\}}(x,y)$ if and only if
$$\suc(x,y) \vee \exists u,v \, (\suc(x,u) \wedge \suc(u,v) \wedge \suc(v,y)) .$$

Another example is the undirected graph $({\mathbb Z}; \Dist_{\{1,2\}})$ with vertex set ${\mathbb Z}$ where two integers $x,y$ are linked in $\Dist_{\{1,2\}}$ if the {\em distance}, $|y-x|$, is one or two.

Structures with a first-order definition in $({\mathbb Z}; \suc)$ are particularly well-behaved from a model-theoretic perspective: all of those structures are strongly minimal~\cite{Marker,Hodges}, and therefore uncountably categorical. Uncountable models of their first-order theory will be
saturated; for implications of those properties for the study of the CSP,
see~\cite{BodHilsMartin}.
In some sense, $({\mathbb Z}; \suc)$ constitutes one of the simplest
infinite structures that is not $\omega$-categorical. 

The corresponding class of CSPs contains many natural
combinatorial problems. For instance, the CSP for the structure
$({\mathbb Z}; \Diff_{\{1,3\}})$ is the computational problem  of
labeling the vertices of a given finite directed graph $G$ such that if $(x,y)$
is an arc in $G$, then the difference between the label for $y$
and the label for $x$ is one or three. It follows from our general results that this problem is in P.
The CSP for the undirected graph 
$({\mathbb Z}; \Dist_{\{1,2\}})$ is exactly the $3$-coloring problem, and thus NP-complete. This is readily seen if one observes 
that any homomorphism of a graph $G$ into the template modulo $3$ gives rise to a $3$-coloring of $G$.
In general, the problems that we study in this paper 
have the flavor of assignment problems where we have to
assign integers to variables such that various given constraints on differences and distances (and Boolean combinations thereof) between variables are satisfied.
We therefore call the class of CSPs whose template is locally finite and definable over $({\mathbb Z}; \suc)$ 
\emph{distance CSPs}. Our main result is the following classification result for distance CSPs.

\begin{Theorem}\label{thm:main}
Let $\Gamma$ be a locally finite structure with a first-order 
definition in $({\mathbb Z};\suc)$. Then at least one of the following applies. 
\begin{itemize}
\item $\Gamma$ has an endomorphism with finite range, and the CSP for $\Gamma$ equals the CSP for a finite structure;
\item the CSP for $\Gamma$ is NP-complete;
\item $\Gamma$ is homomorphically equivalent 
to a structure with a first-order definition in $({\mathbb Z};\suc)$ which has a binary modular max or modular min polymorphism, and the CSP for $\Gamma$ 
is in P.
\end{itemize}
\end{Theorem}

If a locally
finite structure $\Gamma$ with a first-order definition in $({\mathbb Z};\suc)$ 
has a finite core, then a widely accepted conjecture about 
finite domain CSPs implies that the CSP for $\Gamma$ is either NP-complete or in P.
In fact, for this we only need the (open) special case of the conjecture of Feder
and Vardi~\cite{FederVardi} that states that the CSP for finite templates with a transitive automorphism group 
is either in P or NP-complete (see Section~\ref{sect:concl} for details).

To show our theorem, we prove
that if the first two items of the statement do not apply, then
$\Gamma$ is homomorphically equivalent to
a structure $\Delta$ with a first-order definition in $(\Z;\suc)$ that has one of two specific classes of polymorphism which we call \emph{modular
max} and \emph{modular min} (defined in
Section~\ref{sect:modular}). Using these polymorphisms, we further
show that the CSP for $\Delta$, and hence also that for $\Gamma$, can
be solved in polynomial time by certain arc consistency techniques.
Polynomial-time tractability results based on arc consistency
were previously known for finite or $\omega$-categorical
templates; using the local finiteness assumption we manage to apply such techniques to templates which are not $\omega$-categorical.


On the way to our classification result we derive several facts about structures definable in $({\mathbb Z}; \suc)$, and automorphisms and endomorphisms of these structures, which might be of independent interest in model theory, universal algebra, and combinatorics.
For example, we show that every injective endomorphism of
a connected locally finite structure $\Gamma$ with a first-order definition
in $({\mathbb Z}; \suc)$ is either of the form
$x \mapsto -x + c$ or of the form $x \mapsto x + c$ for some $c \in \mathbb Z$ (see Theorem~\ref{thm:endos}).

\section{Preliminaries}\label{sect:prelims}
A \emph{relational signature} $\tau$ is a set of relation
symbols $R_i$, each of which has an associated arity $k_i$. A $\tau$-structure $\Gamma$ consists of a set $D$ (the domain)
together with a relation $R_i^\Gamma \subseteq D^{k_i}$ for each
relation symbol $R_i$ from $\tau$. We consider only finite signatures in this paper.

For $x,y \in \mathbb Z$, let $d(x,y)$ be the distance between
$x$ and $y$, that is, $|y-x|$. The relation $\{(x,y) \; | \; y = x+1\}$
is denoted by $\suc$, and the relation $\{(x,y) \; | \; d(x,y) = 1\}$
is denoted by $\symsuc$. 
It will be convenient to represent 
binary relations $R \subseteq {\mathbb Z}^2$ with
a first-order definition in $(\mathbb Z; \suc)$ 
by sets $S$ of integers
as follows. 
\begin{align*}
\Diff_S := & \{(x,x+k) \; | \; k \in S\} \\
\Dist_S := & \{(x,x+k) \; | \; |k| \in S\}
\end{align*}

A $k$-ary relation $R$ is said to be \emph{first-order (fo)
definable} in a $\tau$-structure $\Gamma$ if there is a first-order $\tau$-formula $\phi(x_1,\ldots,x_k)$ such that $R=\{ (a_1,\ldots,a_{k}) \in D^k \,|\, \Gamma \models \phi(a_1,\ldots,a_{k})\}$. A structure $\Delta$ is said to be fo-definable in $\Gamma$ if $\Delta$ has the same domain as $\Gamma$, and each of its relations is fo-definable in $\Gamma$. For example, $({\mathbb Z}; \symsuc)$
is fo-definable in $({\mathbb Z}; \suc)$ (though the converse is false).

The structure induced by a subset $S$ of the domain of $\Gamma$ is denoted by $\Gamma[S]$. 
When $\Delta_1$ and $\Delta_2$ are two $\tau$-structures
with disjoint domains $D_1$ and $D_2$, then the
\emph{disjoint union} of $\Delta_1$ and $\Delta_2$ is
the structure $\Gamma$ with domain $D_1 \cup D_2$ where $R^{\Gamma} = R^{\Delta_1} \cup R^{\Delta_2}$ for each $R \in \tau$. 
We
say that a structure is \emph{connected} if
it cannot be written as the disjoint union of two non-empty structures.
The \emph{Gaifman graph} of a relational structure $\Gamma$
with domain $D$ is the following undirected reflexive graph: the vertex set is
$D$, and 
there is an edge between elements $x,y \in D$ when $x=y$ or there is a tuple in one of the relations of $\Gamma$ that has both $x$
and $y$ as entries. A structure $\Gamma$ is readily seen to be connected if and only if its Gaifman
graph is connected. The \emph{degree} of a structure
$\Gamma$
is defined to be the degree of the Gaifman graph of $\Gamma$.
The degree of a relation $R \subseteq {\mathbb Z}^k$
is defined to be the degree of the structure $({\mathbb Z}; R)$.
The notation $(\Gamma,R)$ indicates the expansion of $\Gamma$ with the new relation $R$.

A first-order formula is \emph{primitive positive (pp)} if it is of the form 
$$\exists x_1,\dots,x_n \, (\psi_1 \wedge \dots \wedge \psi_m)$$
where $\psi_i$ is an atomic formula over $\Gamma$,
i.e., a formula of the form $y_1=y_2$ or of the form $R(y_1,\dots,y_j)$ for a relation symbol 
$R$ of a relation from $\Gamma$. A pp-sentence is a pp-formula with no free variables. 
For a structure $\Gamma$ with a finite relational signature,
$\mathrm{CSP}(\Gamma)$ is the computational problem of deciding whether
a given pp-sentence is true in $\Gamma$. 
It is not hard to see that
$\mathrm{CSP}(\Delta) \leq_{\mathrm{P}} \mathrm{CSP}(\Gamma)$ 
for any $\Gamma$ and $\Delta$ 
with the same domain such that each of the relations of 
$\Delta$ is pp-definable in $\Gamma$
(see~\cite{JeavonsClosure}); here, $\leq_{\mathrm{P}}$ indicates
polynomial-time many-to-one reduction (though in fact, logspace reductions
may be used).

Suppose $\Gamma$ is a finite structure with finite relational signature $\tau$ and 
domain $D:=\{a_1,\ldots,a_s\}$. Let $\theta_\Gamma(x_1,\ldots,x_s)$ be the
conjunction of the positive facts of $\Gamma$, where the variables
$x_1,\ldots,x_s$ correspond to the elements $a_1,\ldots,a_s$. That is,
$R(x_{\lambda_1},\ldots,x_{\lambda_k})$ appears as an atom in $\theta_\Gamma$
iff $(a_{\lambda_1},\ldots,a_{\lambda_k}) \in R^\Gamma$. Define the
pp-sentence
$\exists x_{1} \ldots x_s. \, \theta_\Gamma(x_1,\ldots,x_s)$ to be the
\emph{canonical query} of $\Gamma$. Conversely, for a pp-sentence $\Theta:=\exists x_{1} \ldots x_s. \, \theta(x_1,\ldots,x_s)$ over the relational signature $\tau$ we define the \emph{canonical database} $\Gamma_{\Theta}$ as follows.
Consider the undirected graph with vertices $x_1,\dots,x_s$ where two vertices $x_i,x_j$ are connected if $\theta$ contains the conjunct $x_i = x_j$. The 
domain of the canonical database is the set of connected components of this graph, and $(C_1,\dots,C_k) \in R$ for
$R \in \tau$ iff there are $y_1 \in C_1,\dots,y_k \in C_k$
such that $\theta$ has a conjunct $\theta(x_1,\ldots,x_s)$.

Let $\Gamma$ and $\Delta$ be $\tau$-structures.  A \emph{homomorphism} from
$\Gamma$ to $\Delta$ is a function $f$ from the domain of $\Gamma$ to the
domain of $\Delta$
such that, for each $k$-ary relation symbol $R$ in $\tau$ and each $k$-tuple
$(a_1, \dots, a_k)$ from $\Gamma$, if $(a_1, \dots, a_k) \in
R^\Gamma$, then $(f(a_1),
\dots, f(a_k)) \in R^\Delta$. In this case we say that the map $f$
\emph{preserves} the relation $R$.
Injective homomorphisms that also preserve
the complement of each relation are called \emph{embeddings}.  Surjective
embeddings are called isomorphisms; homomorphisms and isomorphisms
from $\Gamma$ to itself are called \emph{endomorphisms} and
\emph{automorphisms}, respectively. The set of automorphisms of a
structure $\Gamma$ forms a group under composition. A ($k$-ary)
\emph{polymorphism} of a structure $\Gamma$ over domain $D$ is a
function $f \colon D^k\rightarrow D$ such that, for all $m$-ary relations $R$
of $\Gamma$, if $(a^i_1,\ldots,a^i_m) \in R^\Gamma$, for all $i \leq
k$, then $(f(a^1_1,\ldots,a^k_1), \ldots,f(a^1_m,\ldots,a^k_m) )\in
R^\Gamma$.

A unary function $g$ over domain $D$ is in the \emph{local closure}
of a set of unary functions $F$ over domain $D$ if for every finite
$D' \subseteq D$ there is a function $f' \in F$ such that $g$ and $f'$
agree on all elements in $D'$. We say that $F$ \emph{generates} $f$ if
$f$ is in the local closure of the set $F'$ of all
functions that can be obtained from the members of $F$ by repeated
applications of composition.
It is well-known and easy to see that functions that are
in the local closure of, or
generated by, the endomorphisms of a structure $\Gamma$
are again endomorphisms of $\Gamma$. 


If there exist homomorphisms $f \colon \Gamma \rightarrow \Delta$ and
$g \colon \Delta \rightarrow \Gamma$ then $\Gamma$ and $\Delta$ are said to
be \emph{homomorphically equivalent}. It is a basic observation that CSP$(\Gamma)$ $=$ CSP$(\Delta)$ if $\Gamma$ and $\Delta$ are homomorphically equivalent. A structure is a \emph{core} if
all of its endomorphisms are embeddings \cite{Cores-journal} -- a
\emph{core $\Delta$ of a structure} $\Gamma$ is an induced
substructure that is itself a core and is homomorphically equivalent
to $\Gamma$. It is well-known that if a structure has a finite core,
then that core is unique up to isomorphism (the same is in general not true for
infinite cores).

We could have equivalently 
defined the class of distance
CSPs as the class of CSPs whose template is locally finite and first-order
definable in $({\mathbb Z};s)$, where $s$ is the unary successor \emph{function}, since $({\mathbb Z}; \suc)$ 
and $({\mathbb Z}; s)$ fo-define the same structures.
The structure $({\mathbb Z};s)$ admits \emph{quantifier
elimination}; that is, for every fo-formula $\phi(\overline{x})$ there
is a quantifier-free (qf) $\phi'(\overline{x})$ (possibly equal to \emph{true} or \emph{false}) such that
$({\mathbb Z};s)$ $ \models \forall
{\overline{x}} (\phi(\overline{x}) \leftrightarrow
\phi'(\overline{x}))$; this is easy to prove, and can be found explicitly in~\cite{ModelTheoryShawn}. Thus we may have atomic formulas in $\phi'$ of the form
$y=s^j(x)$, where $s^j$ is the successor function composed on
itself $j$ times. Let $\Gamma$ be a finite signature structure,
fo-definable in $(\mathbb{Z};\suc)$, i.e., qf-definable in its
functional variant $({\mathbb Z};s)$. Let $m$ be the largest number such that
$y=s^m(x)$ appears as a term in the qf definition of a relation of
$\Gamma$. Consider now CSP$(\Gamma)$, the problem of evaluating $\Phi:=
\exists x_1,\ldots,x_k.\phi(x_1,\ldots,x_k)$, where $\phi$ is a
conjunction of atoms, on $\Gamma$. Let $S:=\{1,\ldots,k\cdot (m+1)\}$.
It is not hard to see that $\Gamma \models \mbox{$\Phi$}$ iff
$\Gamma \mbox{$[S]$} \models \mbox{$\Phi$}$. It follows that CSP$(\Gamma)$
will always be in NP.

\paragraph{Convention.} From now on we assume that $\Gamma$ is a relational structure with 
domain $\mathbb Z$ which is first-order definable over
$(\Z;\suc)$ and is locally finite.

\section{Endomorphisms}
\label{sect:endos}
The main result of this section is the following theorem.

\begin{Theorem}\label{thm:endos}
    Let $\Gamma$ be connected. Then:
     \begin{itemize}
        \item $\Gamma$ has either the same automorphisms as $(\Z;\suc)$, or the same automorphisms as $(\Z;\symsuc)$.
        \item Either $\Gamma$ has a finite range endomorphism, or it has an endomorphism 
        whose range induces in $\Gamma$ a structure
        that is isomorphic to a structure which is fo-definable in $({\mathbb Z};\suc)$ and all of whose endomorphisms are automorphisms.
     \end{itemize}
\end{Theorem}
The proof of this theorem can be found at the end of this section,
and makes use of a series of lemmata. 

Before beginning the proof, we remark the following. If $\Gamma$ has a first-order definition in $(\Z;\symsuc)$, then it is easy to see that the automorphisms of $(\Z;\symsuc)$ are also automorphisms of $\Gamma$,
and hence the two structures have the same automorphisms by Theorem~\ref{thm:endos}. 
Now it is tempting to believe that also the converse holds, i.e., that if $\Gamma$ has the same automorphisms as $(\Z;\symsuc)$, 
 then $\Gamma$ is fo-definable in $(\Z;\symsuc)$
  (this would be true for $\omega$-categorical structures).
 However, this is not true: Let
$$
    R:=\{(x,y,u,v)\in\Z^4 \,|\, (\suc(x,y) \wedge \suc(u,v))\vee (\suc(v,u) \wedge \suc(y,x))\},
$$
and set $\Gamma:=(\Z;R)$. The function which sends every $x\in\Z$ to $-x$ is an automorphism of $\Gamma$, so the automorphism group of $\Gamma$ equals that of $(\Z;\symsuc)$, by Theorem~\ref{thm:endos}. However, $R$ is not fo-definable in $(\Z;\symsuc)$. To see this, suppose it were definable. Then $R$ is also definable in $(\Z; \Dist_{\{1\}}, \Dist_{\{2\}}, \ldots)$, and even with a quantifier-free formula $\phi(x,y,u,v)$ since this structure has quantifier-elimination. Let $n$ be the maximal natural number such that $\Dist_{\{n\}}$ occurs in $\phi(x,y,u,v)$. We claim that $\phi(0,1,n+2,n+3)$ holds iff $\phi(0,1,n+3,n+2)$ holds. To see this, we show that any atom of the formula $\phi(x,y,u,v)$, i.e., any occurrence of $\Dist_{\{k\}}(a,b)$, where $\{a,b\}\subseteq\{x,y,u,v\}$ and $k\leq n$, evaluates to \emph{true} upon insertion of $v_1:=(0,1,n+2,n+3)$ for the variables $(x,y,u,v)$ if and only if it evaluates to \emph{true} upon insertion of $v_2:=(0,1,n+3,n+2)$ for $(x,y,u,v)$. This is obvious when $\{a,b\}\subseteq\{x,y\}$ since $v_1$ and $v_2$ have identical values for $x,y$. If $|\{a,b\}\cap\{x,y\}|=1$ then the atom becomes \emph{false} in both evaluations, so the only remaining case is where $\{a,b\}\subseteq\{u,v\}$; but then the atom becomes true in both evaluations if and only if $k=1$ and $a\neq b$, so we are done. Now since $\phi(0,1,n+2,n+3)$ holds iff $\phi(0,1,n+3,n+2)$ holds, we have a contradiction since $v_1$ is an element of $R$ whereas $v_2$ is not.

Denote by $E$ the edge-relation of the Gaifman graph of $\Gamma$. It is clear that every endomorphism of $\Gamma$ preserves $E$. We claim that there are $0<d_1<\cdots < d_{n}$ such that $E(x,y)$ holds iff $d(x,y)\in\{0,d_1,\ldots,d_{n}\}$. To see this, observe that if $(x,y) \in E$ and $u,v\in \Z$ are so that $d(x,y)=d(u,v)$, then also $(u,v) \in E$, because there is an automorphism of $(\Z;\suc)$ (and hence of $\Gamma$) which sends $\{x,y\}$ to $\{u,v\}$ and this automorphism also preserves $E$. Hence, the relation $E$ is determined by distances. Moreover, there are only finitely many distances since $\Gamma$ is assumed to have finite degree.

\begin{Notation}\label{notation:distances}
    We will refer to the distances defining the Gaifman graph of $\Gamma$ as $d_1,\ldots,d_n$. We also write $D$ for the largest distance $d_n$.
\end{Notation}

The following basic claim characterizes when $\Gamma$ is connected in terms of the distance set.

\begin{Lemma}
\label{lem:4}
$\Gamma$ is connected if and only if the greatest common divisor of $d_1,\ldots,d_n$ is $1$.
\end{Lemma}
\begin{Proof}
If $d$ is the greatest common divisor of $d_1,\ldots,d_n$ it is clear that all the nodes accessible from a node $x\in\Z$ are of the form $x+c\cdot d$ where $c\in\Z$.
Conversely, every node of the form $x+c\cdot d$ is accessible from $x$ because
$c\cdot d=c_1\cdot d_1+\dots+c_n\cdot d_n$ for some $c_1,\dots,c_n\in \Z$,
by the extended Euclidean algorithm.
\end{Proof}

In order to lighten the notation we might use $ex$ to denote $e(x)$, where $e$ is an endomorphism of $\Gamma$ and $x\in \Z$.

\begin{Lemma}\label{lem:orbit-bound}
    Suppose that $\Gamma$ is connected. Then there exists a constant $c=c(\Gamma)$ such that for all endomorphisms $e$ of $\Gamma$ we have $d(e(x),e(y))\leq d(x,y)+ c$ for all $x,y\in\Z$.
\end{Lemma}
\begin{Proof}
    We first claim that for every $0<q<D$, there exists a number $c_q$ such that $d(e(x),e(y))\leq c_q$ for all endomorphisms $e$ of $\Gamma$ and all $x,y\in\Z$ with $d(x,y)=q$. To see this, pick $u,v$ with $d(u,v)=q$ and a path between $u$ and $v$ in the Gaifman graph of $\Gamma$; say this path has length $l_q$. Then, since this path is mapped to a path under any endomorphism, we have $d(e(u),e(v))\leq D\cdot l_q$ for all endomorphisms $e$. Since an isomorphic path exists for all $x,y$ with the same distance, our claim follows by setting $c_q:=D\cdot l_q$. Set $c$ to be the maximum of the $c_q$, and let an endomorphism $e$ and $x,y\in\Z$ be given. Assume without loss of generality that $x<y$. There exists $m\geq 0$ and $0\leq q<D$ such that $y=x+D\cdot m+q$. Set $x_r:=x+D\cdot r$, for all $0\leq r\leq m$. Since $x_r$ and $x_{r+1}$ are adjacent in the Gaifman graph of $\Gamma$  for all $0\leq r< m$, so are $ex_r$ and $ex_{r+1}$, and hence $d(ex_r,ex_{r+1})\leq D$. Therefore,
\begin{eqnarray*}
\begin{aligned}
    d(ex,ey)&\leq \sum_{0\leq r< m} d(ex_r,ex_{r+1})+d(ex_m,ey)\leq
    D\cdot m+ d(ex_m,ey)\\
    &\leq d(x,y)+d(ex_m,ey)\leq d(x,y)+ c.
\end{aligned}
\end{eqnarray*}
\end{Proof}

Observe that a constant $c(\Gamma)$ not only exists, but can actually be calculated given the distances $d_1,\ldots,d_n$: by the proof of Lemma~\ref{lem:orbit-bound}, it suffices to calculate a constant $c_q$ for all $0<q<D$. To do this, one must find a path of length $l_q$ between two numbers $u,v\in\Z$ with $d(u,v)=q$; this again amounts to solving the equation $x_1\cdot d_1 +\cdots+ x_n\cdot d_n=q$ (with variables $x_1,\ldots,x_n$) over $\Z$, which can be achieved by the extended Euclidean algorithm.

In the following, we will keep the symbol $c$ reserved for the minimal constant guaranteed by the preceding lemma.

\begin{Lemma}\label{lem:finiterange}
 Suppose that $\Gamma$ is connected, and let $e$ be 
 an endomorphism of $\Gamma$ with the property
 that for all $k>c+1$ there exist $x,y \in \Z$ with $d(x,y)=k$ and $d(e(x),e(y))<k$. 
Then $\Aut({\mathbb Z};\suc) \cup \{e\}$ generates
an endomorphism whose range has size at most $2(c+1)$. 
\end{Lemma}
\begin{Proof}
Let $A\subseteq \mathbb{Z}$ be finite. We claim that $F:=\Aut({\mathbb Z};\suc) \cup \{e\}$ generates a
function $f_A$ which maps $A$ into a set of diameter at most $2c + 1$.
The lemma then follows by the following standard local closure
argument: Let $S$ be the set of all those functions $\alpha$ whose
domain is a finite interval $[-n;n]\subseteq \Z$ and whose range is
contained in the interval $[-c;c]$, and which have the property that
there exists a function generated by $F$ which agrees with $\alpha$ on
$[-n;n]$. By our claim, $S$ is infinite. For functions $\alpha, \beta$
in $S$, write $\alpha\leq \beta$ iff $\beta$ is an extension of
$\alpha$. Clearly, the set $S$, equipped with this order, forms a
finitely branching tree; since the tree is infinite, it has an
infinite branch (this easily verified fact is called K\"{o}nig's
lemma) $B\subseteq S$. The branch $B$ defines a function $f$ from $\Z$
into the interval $[-c;c]$; since $F$ generates functions which agree
with $f$ on arbitrarily large intervals of the form $[-n;n]$, we have
that $f$ is generated by $F$, too. This completes the proof.

    Enumerate the pairs $(x,y)\in A^2$ with $x<y$ by $(x_1,y_1),\ldots,(x_r,y_r)$. Now the hypothesis of the lemma implies that by successive applications of $e$ and shifts we can map $(x_1,y_1)$ to a pair of distance at most $c+1$; in other words, there exists $t_1$ generated by $F$ such that $d(t_1x_1,t_1y_1)\leq c+1$. Similarly, there exists $t_{2}$ generated by $F$ such that $d(t_{2}t_1 x_{2},t_{2}t_1 y_{2})\leq c+1$. Continuing like this we arrive at a function $t_{r}$ generated by $F$ such that $d(t_{r}t_{r-1}\cdots t_1 x_{r},t_{r}t_{r-1}\cdots t_1 y_{r})\leq c+1$. Now consider $t:=t_r\circ\cdots \circ t_1$. Set $f_j:=t_{r}\circ \cdots\circ t_{j+1}$ and $g_j:=t_{j}\circ\cdots\circ t_{1}$, for all $1\leq j\leq r$; so $t=f_j\circ g_j$. Then, since by construction $d(g_j(x_j),g_j(y_j))\leq c+1$, we have that for all $1\leq j\leq r$
    \begin{align*}
    d(tx_j,ty_j) = & \; d(f_j(g_j(x_j)),f_j(g_j(y_j)) \\
    \leq & \; d(g_j(x_j),g_j(y_j))+c && \text{(Lemma~\ref{lem:orbit-bound})} \\
    \leq & \; 2 c+1
    \end{align*} 
    and our claim follows.
\end{Proof}

\begin{Lemma}\label{lem:periodic}
 Suppose that $\Gamma$ is connected with an endomorphism $e$ that does not
 satisfy the hypothesis of the preceding lemma, i.e., 
 there exists $k> c+1$ such that $d(ex,ey)\geq k$ for all $x,y$ with $d(x,y)=k$. Then either $e(s+D)=e(s)+D$ for all $s\in {\mathbb Z}$ or $e(s+D)=e(s)-D$ for all $s\in{\mathbb Z}$. 
\end{Lemma}

\begin{Proof}
Let $k>c+1$ be so that $d(ex,ey)\geq k$ for all $x,y$ with $d(x,y)=k$. Let $w\in \mathbb{Z}$ be arbitrary. Then, since $d(e(w+k),e(w))\geq k$, we have $e(w)\neq e(w+k)$. We furthermore assume that $e(w+k)>e(w)$; the situation where $e(w+k)<e(w)$ can be treated symmetrically. 
We claim that $e(v+k)\geq e(v)+k$ for all $v\in \mathbb{Z}$. Suppose not, and say without loss of generality that there exists $v>w$ contradicting our claim. Then, since $d(e(v+k),e(v))\geq k$, we have $e(v+k)\leq e(v)-k$. Take the minimal $v$ with $v>w$ satisfying this property. Then, by minimality, we have $e(v-1+k)\geq e(v-1)+k$. Since by Lemma~\ref{lem:orbit-bound} we have $d(e(v-1+k),e(v+k))\leq c+1$, we get that $e(v-1)+k-c-1\leq e(v+k)$. On the other hand, $e(v)-c-1\leq e(v-1)$. Inserting this into the previous inequality, we obtain $e(v)-c-1+k-c-1\leq e(v+k)$, which yields  $e(v)-2c-2+k\leq e(v+k)$. By our assumption on $v$, we obtain $e(v)-2c-2+k\leq e(v)-k$, which yields $k\leq c+1$, a contradiction.

Set $b:=k\cdot D$. We next claim that $e(v+b)= e(v)+b$ for all $v\in{\mathbb Z}$. First observe that points at distance $D$ cannot be mapped by $e$ to points at larger distance since $D$ is by definition the largest distance in the Gaifman graph of $\Gamma$. 
Since $b$ is a multiple of $D$, we get that $e(v+b)\leq e(v)+b$. On the other hand, since $b$ is also a multiple of $k$ and since $e(v+k)\geq e(v)+k$ for all $v\in\Z$, we obtain $e(v+b)\geq e(v)+b$, proving the claim.

We now prove that $e(v)+D\leq e(v+D)$ for all $v\in{\mathbb Z}$. This is because 
\begin{align*}
e(v)+kD = & \; e(v)+b = e(v+b) = e(v+kD) \\
= & \; e(v+D+ (k-1)D) \\
\leq & \; e(v+D)+(k-1)D
\end{align*} 
the latter inequality holding since $D$ is the maximal distance in the relation $E$ and cannot be increased. Subtracting $(k-1)D$ on both sides, our claim follows.

Since points at distance $D$ cannot be mapped to points at larger distance under $e$, we have $e(v+D)\leq e(v)+D$ for all $v\in\Z$, and we have proved the lemma.
\end{Proof}

The following lemma summarizes the preceding two lemmas.

\begin{Lemma}\label{lem:summaryPeriodic}
 Suppose that $\Gamma$ is connected. 
    The following are equivalent for an endomorphism $e$ of $\Gamma$:
    \begin{itemize}
        \item[(i)] There exists $k>c+1$ such that $d(ex,ey)\geq k$ for all $x,y\in\Z$ with $d(x,y)=k$.
        \item[(ii)] $\Aut(\Z;\suc) \cup \{e\}$ does not generate a finite range operation.
        \item[(iii)] $e$ satisfies either $e(v+D)=e(v)+D$ for all $v \in \Z$, or $e(v+D)=e(v)-D$ for all $v \in \Z$.
    \end{itemize}
\end{Lemma}
\begin{Proof}
    Lemma~\ref{lem:periodic} shows that (i) implies (ii) and (iii). It follows from Lemma~\ref{lem:finiterange} that (ii) implies (i). Finally, it is clear that (iii) implies (ii).
\end{Proof}

We know now that there are two types of endomorphisms of $\Gamma$: Those which are periodic with period $D$, and those which generate a finite range operation. We will next provide examples showing that both types really occur.

\begin{Example}\label{ex:endo1}
     Set $\Gamma:=(\Z;\Dist_{\{1,3\}})$.
     Set $e(3k):=3k$, $e(3k+1):=3k+1$, and $e(3k+2):=3k$, for all $k\in\Z$. Then $e$ is an endomorphism of $\Gamma$ that does not generate any finite range operations since it satisfies $e(v+3)=e(v)+3$ for all $v\in \Z$.
\end{Example}

Observe that in the previous example, we checked that $e$ is of the non-finite-range type by virtue of the easily verifiable Item (iii) of Lemma~\ref{lem:summaryPeriodic} and without calculating $c(\Gamma)$, which would be more complicated.

\begin{Example}
     For the structure $\Gamma$ from Example~\ref{ex:endo1}, let $e$ be the function which maps every $x\in\Z$ to its value modulo $4$. Then $e$ is an endomorphism which has finite range.
\end{Example}

\begin{Example}
    The structure $\Gamma:=(\Z;\Dist_{\{1,3,6\}},\Dist_{\{3\}})$ has the endomorphism from Example~\ref{ex:endo1}.
    However, it does not have any finite range endomorphism. To see this, consider the set $3 \Z:=\{3m \, | \, m\in\Z\}$. If $e$ were a finite range endomorphism, it would have to map this set onto a finite set. By composing $e$ with automorphisms of $(\Z; \suc)$, we may assume that $e(0)=0$ and $e(3)>0$. Then $e(3)=3$ as $e$ preserves $\Dist_{\{3\}}$.  We claim $e(s)=s$ for all $s\in 3\Z$. Suppose to the contrary that $s$ is the minimal positive counterexample (the negative case is similar). We have $e(s-3)=s-3$ and hence $e(s)\in\{s-6,s\}$ because $e$ preserves $\Dist_{\{3\}}$. If we had $e(s)=s-6$, then $e(s-6)=s-6$ and $(s-6,s)\in \Dist_{\{1,3,6\}}$ yields a  contradiction.
\end{Example}

\begin{Example}
    Let $\Gamma=(\Z;\symsuc)$, and let $e$ be the function that maps every $x$ to its absolute value. Then $e$ does not have finite range, but generates with $\Aut(\Z;\suc)$ a function with finite range (namely, the function which sends the even numbers to $0$ and the odd numbers to $1$).
\end{Example}

The proof of Lemma~\ref{lem:periodic} generalizes canonically to a more general situation.

\begin{Lemma}
Suppose that $\Gamma$ is connected.
    Let $e$ be an endomorphism of $\Gamma$ satisfying the various statements of Lemma~\ref{lem:summaryPeriodic}. Let $q$ be so that $d(x,y)=q$ implies that $d(ex,ey)\leq q$. Then $e$ satisfies either $e(v+q)=e(v)+q$ for all $v\in\Z$, or $e(v+q)=e(v)-q$ for all $v\in\Z$.
\label{lemma:13}
\end{Lemma}
\begin{Proof}
    This is the same argument as in the proof of Lemma~\ref{lem:periodic}, with $D$ replaced by $q$.
\end{Proof}

\begin{Definition}
    Given an endomorphism $e$ of $\Gamma$, we call all positive integers $q$ with the property that $e(v+q)=e(v)+q$ for all $v\in\Z$ or $e(v+q)=e(v)-q$ for all $v\in\Z$ \emph{stable for $e$}.
\end{Definition}

Observe that if $e$ satisfies the various statements of Lemma~\ref{lem:summaryPeriodic}, then $D$ is stable for $e$. Note also that if $p,q$ are stable for $e$, then they must have the same ``direction'': We cannot have $e(v+p)=e(v)+p$ and $e(v+q)=e(v)-q$ for all $v\in\Z$.

\begin{Lemma}
Suppose that $\Gamma$ is connected.
    Let $e$ satisfy the various statements of Lemma~\ref{lem:summaryPeriodic}, and let $q$ be the minimal stable number for $e$. Then the stable numbers for $e$ are precisely the multiples of $q$. In particular, $q$ divides $D$.
\end{Lemma}
\begin{Proof}
    Clearly, all multiples of $q$ are stable. Now for the other direction suppose that $p$ is stable but not divisible by $q$. Write $p=m\cdot q+r$, where $m,r$ are positive numbers and $0< r<q$. Since $r$ is not stable, composing $e$ and shifts we can build a function $t$ such that $t(0)=0$ and $d(t(mq),t(p))\neq r$. By the property of $p$ we should have $t(p)=p$ or  $t(p)=-p$. But this is impossible since then $d(t(mq),t(p))=d(mq,p)=r$, a contradiction.
\end{Proof}

\begin{Lemma}\label{lem:infiniteCore}
Suppose that $\Gamma$ is connected and has 
an endomorphism $e$ satisfying the statements of Lemma~\ref{lem:summaryPeriodic}. Let $q$ be its minimal stable number.
    Then there is an endomorphism $t$ of $\Gamma$ which can be written as a functional composite using automorphisms of $({\mathbb Z};\suc)$ and $e$ which has the following properties:
     \begin{itemize}
        \item $t$ satisfies either $t(v+q)=t(v)+q$ or $t(v+q)=t(v)-q$
        \item $t(0)=0$
        \item $t[{\mathbb Z}]=\{q\cdot z \, | \, z\in{\mathbb Z}\}$.
     \end{itemize}
\end{Lemma}

\begin{Proof}
Assume $1<q$ (otherwise $t$ can be chosen to be the identity and there is nothing to do). We claim that 
$\Aut({\mathbb Z};\suc) \cup \{e\}$ generates a function $t_1$ such that $t_1(0)=0$ and $t_1(1)\in\{q\cdot z \, | \, z\in{\mathbb Z}\}$. To see this, observe that since $1<q$ and since $q$ is the smallest positive number with the property that $d(x,y)=q$ implies $d(ex,ey)\leq q$ (Lemma~\ref{lemma:13}), there exist $x_0,y_0\in{\mathbb Z}$ with $d(x_0,y_0)=1$ and $d(ex_0,ey_0)>1$. Write $r_1:=d(ex_0,ey_0)$. If $r_1$ is not a multiple of $q$, then there exist $x_1,y_1\in\Z$ with $d(x_1,y_1)=r_1$ and $d(ex_1,ey_1)=:r_2>r_1$. Again, if $r_2$ is not a multiple of $q$, then there exist $x_2,y_2\in\Z$ with $d(x_2,y_2)=r_2$ and $d(ex_2,ey_2)=:r_3>r_2$. Consider the sequence $(x_i,y_i)$ of pairs of distance $r_i$ (setting $r_0:=1$). By exchanging $x_{i+1}$ and $y_{i+1}$ if necessary, we may assume that $x_{i+1}<y_{i+1}$ iff $ex_i< ey_i$, for all $i$. There exist automorphisms $\alpha_i$ of $({\mathbb Z};\suc)$ such that $(\alpha_i(e(x_i)),\alpha_i(e(y_i)))=(x_{i+1},y_{i+1})$. Set $s_i:=\alpha_i\circ e\circ \alpha_{i-1}\circ\cdots\circ \alpha_0\circ e$. Then the endomorphism $s_i$ sends $(x_0,y_0)$ to $(x_{i+1},y_{i+1})$, a pair of distance $r_{i+1}>r_i>\cdots>r_0$. Thus the sequence must end at some finite $i$, by Lemma~\ref{lem:orbit-bound}. By construction of the sequence, this happens only if $r_{i+1}$ is a multiple of $q$. Therefore, $r_{i+1}=d(s_i(x_0),s_i(y_0))\in\{q\cdot z \, | \, z\in{\mathbb Z}\}$. By applying shifts we may assume $x_0=0$, $y_0=1$, and $s_i(0)=0$. Set $t_1:=s_i$.

Now if $2<q$, then consider the number $t_1(2)$. We claim that $\Aut({\mathbb Z};\suc) \cup \{e\}$ generates a function $t_2$ such that $t_2(0)=0$ and $t_2(t_1(2))$ is a multiple of $q$. If already $t_1(2)$ is a multiple of $q$, then we can choose $t_2$ to be the identity. Otherwise, we can increase the distance of $t_1(2)$ from $0$ successively by applying shifts and $e$ just as before, where we moved away $1$ from $0$. After a finite number of steps, we arrive at a function $t_2$ such that $d(t_2(0),t_2t_1(2))$ is a multiple of $q$. Applying a shift one more time, we may assume that $t_2(0)=0$, and so $t_2$ has the desired properties.

We continue inductively, constructing for every $i<q$ a function $t_i$ such that $t_i(0)=0$ and 
$t_i\circ\cdots \circ t_1(j)$ is a multiple of $q$
for all $j \leq i$. 
At the end, we set $t:=t_{q-1}\circ\cdots\circ t_1$. Since $e$ satisfies either $e(v+q)=e(v)+q$ or $e(v+q)=e(v)-q$, so does
$t$, as it is composed of $e$ and automorphisms of $(\Z; \suc)$. It is
also clear from the construction that $t(0)=0$ holds. These two facts
together imply that $t[\Z]$ contains the set $\{q\cdot z \, | \, z\in{\mathbb
Z}\}$. For the other inclusion, let $v\in \Z$ be arbitrary, and write
$v= q\cdot z+ r$, where $z\in \Z$ and $0\leq r<q$. Then $t(v)=q\cdot z+
t(r)$ or $t(v)= - q\cdot z+ t(r)$, which is a multiple of $q$ since
$t(r)$ is a multiple of $q$ by construction.
\end{Proof}

Observe that we did not need local closure in the preceding lemma.

\begin{Lemma}\label{lem:notInjective}
Suppose that $\Gamma$ is connected and has an endomorphism $e$ which is not an automorphism of $({\mathbb Z};\symsuc)$. Then $e$ is not injective.
\end{Lemma}
\begin{Proof}
    If $\Aut({\mathbb Z};\suc) \cup \{e\}$ generates a finite range operation then the lemma follows immediately, so assume this is not the case. Then $e$ has a minimal stable number $q$. Since $e$ is not an automorphism of $({\mathbb Z};\symsuc)$, we have $q>1$. But now the statement follows from the preceding lemma, since the function $t$ is not injective (e.g., $t$ maps $\{q\cdot z : z \in\mathbb{Z}\}$ surjectively to $\{q\cdot z : z \in\mathbb{Z}\}$, so $t(1)=t(w)$ for some $w \in \{q\cdot z : z \in\mathbb{Z}\}$).
\end{Proof}

\begin{Lemma}
Suppose that $\Gamma$ is connected and has an endomorphism which is not an automorphism of $({\mathbb Z};\symsuc)$ such that $\{e\} \cup \Aut({\mathbb Z};\suc)$ does not generate a finite range operation. Then $e$ is not surjective.
\end{Lemma}
\begin{Proof}
    This is a direct consequence of Lemma~\ref{lem:infiniteCore}, since being surjective is preserved under composition (and we used just composition in Lemma~\ref{lem:infiniteCore} and not local closure).
\end{Proof}

\vspace{0.2cm}
Define $\Gamma/k$ to be the substructure
of $\Gamma$ induced by $\{k\cdot z \, | \, z\in{\mathbb Z}\}$. Note that when $\Gamma$ is fo-definable in $(\Z;\suc)$, then 
$\Gamma/k$ is isomorphic to a structure $\Delta$ that 
is fo-definable in $(\Z;\suc)$ via the map which sends
an element $x$ of $\Gamma/k$ to $x/k \in {\mathbb Z}$. From a defining quantifier-free formula $\phi$ for a relation $R^\Gamma$ of $\Gamma$ over $(\Z;s)$, we obtain a definition 
for $R^\Delta$ over $(\Z;s)$ as follows. 
For all $i\in\omega$ not divisible by $k$, replace every occurrence of $s^i$ by $\forall x (x\neq x)$. For all other $i$, replace every occurrence of $s^i$ by $s^{i/k}$. 

\vspace{.5cm}

\begin{Proof} (of Theorem~\ref{thm:endos})
    We prove the first statement. It is a direct consequence of Lemma~\ref{lem:notInjective} that the automorphism group of $\Gamma$ is contained in that of $(\Z;\symsuc)$. Since $\Gamma$ is fo-definable in $(\Z;\suc)$, its automorphism group contains that of $(\Z;\suc)$. The statement now follows from the easily verifiable fact that there are no permutation groups properly between the automorphism groups of $(\Z;\suc)$ and $(\Z;\symsuc)$.

    For the second statement, suppose that $\Gamma$ has no finite range endomorphism. If all of its endomorphisms are automorphisms, then we are done. Otherwise, $\Gamma$ has an endomorphism $t$ as in Lemma~\ref{lem:infiniteCore}, with $q>1$. 
    Let $\Delta$ be a structure that is isomorphic
    to $\Gamma/q$ and first-order definable in $(\Z;\suc)$. 
   In $\Gamma/q$, 
    two points $x, y$ are adjacent iff $d(x,y)\in\{d_1,\ldots d_n\}$; moreover, $d(x,y)$ is divisible by $q$. Therefore, the remaining relevant distances are those divisible by $q$. In other words, if $\{d_{i_1},\ldots,d_{i_r}\}$ are those distances from $\{d_1,\ldots,d_n\}$ which are divisible by $q$, then the Gaifman graph of $\Delta/q$ is isomorphic to the graph on $\Z$ defined by the distances $\{\frac{d_{i_1}}{q},\ldots,\frac{d_{i_r}}{q}\}$.
    Since before, from Lemma~\ref{lem:4}, the greatest common divisor of all possible distances was $1$, we must have lost at least one distance, i.e., $r<n$.

    Observe that $\Gamma/q$ (and hence $\Delta$) 
    is connected as it is the image of an endomorphism of $\Gamma$. Note moreover that $\Gamma/q$ (and hence $\Delta)$ cannot have a finite range endomorphism: If $s$ were such an endomorphism, then $s\circ t$ would be a finite range endomorphism for $\Gamma$, contrary to our assumption. If all endomorphisms of $\Gamma/q$ are automorphisms, then we are done. Otherwise $\Delta$ satisfies all assumptions that we had on $\Gamma$, and we may repeat the argument. Since in every step we lose a distance for the Gaifman graph, this process must end, meaning that we arrive at a structure all of whose endomorphisms are automorphisms.
\end{Proof}

\section{Definability of Successor}
\label{sect:def-succ}
In this section we show how to reduce the complexity classification
for distance constraint satisfaction problems with template $\Gamma$ to the case where
either $\Gamma$ has a finite core, 
or the relation $\suc$ is pp-definable in $\Gamma$. 
We make essential use of the results of the previous section; 
but note that in this section we do \emph{not}
assume that $\Gamma$ is connected.

\begin{Theorem}\label{thm:def-succ}
Suppose that $\Gamma$ does not have an endomorphism
of finite range. Then 
$\Gamma$ is homomorphically equivalent to 
a connected finite-degree structure $\Delta$ with a
first-order definition in $({\mathbb Z}; \suc)$ which satisfies one of two possibilities:
$\Csp(\Delta)$ (and, hence, $\Csp(\Gamma)$) is NP-hard, or $\suc$ is
definable in $\Delta$.
\end{Theorem}

The following lemma demonstrates how the not necessarily 
connected case can be reduced to the connected case.

\begin{Lemma}\label{lem:disconnected}
$\Gamma$ is homomorphically equivalent to a \emph{connected} finite-degree structure $\Delta$ with a
first order definition in $({\mathbb Z}; \suc)$.
\end{Lemma}
\begin{Proof}
If all edges of the Gaifman graph of $\Gamma$ are self-loops,
then the statement is clear. Otherwise,
let $g$ be the greatest common divisor of $d_1,\dots,d_n$
(the \emph{distances} in the Gaifman graph, see Section~\ref{sect:endos}, Notation~\ref{notation:distances}). 
If $\Gamma$ is connected, there is nothing to prove. 

Otherwise, if $\Gamma$ is disconnected, by 
Lemma~\ref{lem:4},
we have $g>1$.  Then $\Gamma$ must be a disjoint union of $g$ copies 
of a connected structure $\Delta$ (and these copies are isomorphic to
each other by an isomorphism of the form $x \mapsto x+d$, for appropriate constant $d$). 
In particular, $\Gamma$ is homomorphically equivalent to $\Delta$.
Moreover, $\Delta$
itself has a first-order definition in $(\mathbb Z; \suc)$. 
The proof here is as in the proof of Theorem~\ref{thm:endos}, 
with $g$ taking the role of $q$.
\end{Proof} 
\vspace{.4cm}


The following is obvious.

\begin{Lemma}\label{lem:orbits}
Let $(a_1,\dots,a_k),(b_1,\dots,b_k) \in {\mathbb Z}^k$. Then there is an automorphism $\alpha$ of $({\mathbb Z}; \suc)$ with $\alpha(a_i)=b_i$ for all $i \leq k$ if and only if
 $a_i-a_j = b_i-b_j$ for all $1 \leq i,j \leq k$.
\end{Lemma}

\begin{Lemma}\label{lem:connectedness}
Suppose that $\Gamma$ is connected.
Then there is an $n_0$ such that 
the structure $\Gamma[\{1,\dots,n\}]$ is connected for all $n \geq n_0$.
\end{Lemma}
\begin{Proof}
Let $d_1$ be the smallest distance of the
distances $\{d_1,\dots,d_n\}$ defining the Gaifman graph $G$ of $\Gamma$ (as in Section~\ref{sect:endos}).
By connectivity of $G$, for each pair $a,b$ of elements from $\{1,\dots,d_1\}$ there is a path from $a$ to $b$ in $G$. 
Fix such a path for each pair $a,b$. 
Let $n_0$ be the smallest number such that
all vertices on those paths are smaller than $n_0$. 
We claim that $\Gamma[\{1,\dots,n\}]$ is connected for all $n \geq n_0$. To see that $c,d \leq n$ are connected, observe that both $c$ and $d$ are connected to vertices in $\{1,\dots,d_1\}$ (via a sequence
of vertices at distance $d_1$).
Since all vertices in $\{1,\dots,d_1\}$ are connected in $\Gamma[\{1,\dots,n_0\}]$ by construction, we conclude that $c$ and $d$ are connected by a path in
$\Gamma[\{1,\dots,n\}]$.
\end{Proof}


\begin{Lemma}\label{lem:orbit-bound-finite}
Suppose that $\Gamma$ is connected. Then there is an $n_0$ and $c$ such that for all $n \geq n_0$
and any homomorphism $f$ from 
$\Gamma[\{1,\dots,n\}]$ to $\Gamma$ we have
 that $d(f(x),f(y)) \leq c + d(x,y)$ for all $x,y \in \{1,\dots,n\}$.
\end{Lemma}
\begin{Proof}
Let $n_0$ be the number from Lemma~\ref{lem:connectedness}.
Then for all $n \geq n_0$, the structure $\Gamma[\{1,\dots,n\}]$ is connected.
Now, proceed as in Lemma~\ref{lem:orbit-bound}.
\end{Proof}


\begin{Proposition}\label{prop:compactness}
Let $\Gamma$ be connected and such that every endomorphism of $\Gamma$ is an automorphism
of $({\mathbb Z}; \symsuc)$. Then for all $a_1,a_2 \in \mathbb Z$ 
there is a finite $S \subseteq \mathbb Z$ that contains 
$\{a_1,a_2\}$ such that for all homomorphisms $f$ from $\Gamma[S]$ to $\Gamma$ we have $d(f(a_1),f(a_2))=d(a_1,a_2)$.
\end{Proposition}

\begin{Proof}
Suppose that there are $a_1 < a_2 \in \Gamma$ such that for
all finite subsets $S$ 
of elements of $\Gamma$ that contain $\{a_1,a_2\}$
there is a homomorphism from $\Gamma[S]$ to $\Gamma$ where
$d(f(a_1),f(a_2)) \neq d(a_1,a_2)$.
We have to show that $\Gamma$ has an endomorphism that
is not an automorphism of $({\mathbb Z}; \symsuc)$.
Let $S$ be a subset of $\mathbb Z$ that contains $\{a_1,a_2\}$, and let $f,g$ be functions
from $S \rightarrow \mathbb Z$. Then we define $f \sim g$ if there exists an automorphism $\alpha$ of $\Gamma$ such that $f(x)=\alpha(g(x))$ for all $x \in S$. Homomorphisms from $\Gamma[S]$ to $\Gamma$ where $d(f(a_1),f(a_2)) \neq d(a_1,a_2)$ will be called \emph{good}. Observe that since
all automorphisms of $\Gamma$ preserve distances, if one function in an equivalence class is good, then all other
functions in the equivalence class are also good.
 
Let $n_0$ be the number from Lemma~\ref{lem:orbit-bound-finite},
and let $n_1$ be $\max(n_0,|a_1|,|a_2|)$. 
Consider the following infinite forest $\cal T$:
the vertices are the equivalence 
classes of good functions $f \colon  V \rightarrow \mathbb Z$ for $V = \{-n,\dots,n\}$, for all $n \geq n_1$, and $\cal T$ has an arc from one such equivalence class $F$ to another $H$ 
if there are $f \in F$, $h \in H$, such that $f$ is a restriction of $h$,
and $f$ is defined on $\{-n,\dots,n\}$, and $h$ is defined on $\{-n-1,\dots,n+1\}$, for some $n \in \mathbb N$. Observe that
\begin{itemize}
\item by our assumptions the forest $\cal T$ is infinite;
\item by Lemma~\ref{lem:orbit-bound-finite}, for every $n \geq n_1$ there is a $b$ such that 
$d(f(x),f(y))< b $ for all $x,y \in \{-n,\dots,n\}$.  
Using Lemma~\ref{lem:orbits} it follows that $\cal T$ is finitely branching;
\item the forest $\cal T$ has only finitely many roots.
\end{itemize}

By K\"onig's lemma, there is an infinite branch in $\cal T$.
It is straightforward to use this infinite branch to construct an 
endomorphism $f$ of $\Gamma$ with $d(a_1,a_2) \neq
d(f(a_1),f(a_2))$. This endomorphism cannot
be an automorphism of $({\mathbb Z}; \symsuc)$, which
 concludes the proof.
\end{Proof}

\begin{Proposition}\label{prop:compactness2}
Let $\Gamma$ be connected and such that every endomorphism of $\Gamma$ is an automorphism
of $({\mathbb Z}; \suc)$. Then for all $a_1,a_2 \in \mathbb Z$ 
there is a finite $S \subseteq \mathbb Z$ that contains 
$\{a_1,a_2\}$ such that for all homomorphisms $f$ from $\Gamma[S]$ to $\Gamma$ we have $f(a_1)-f(a_2)=a_1-a_2$.
\end{Proposition}
\begin{Proof}
The proof is similar to the proof of Proposition~\ref{prop:compactness}. 
\end{Proof}


\begin{Corollary}\label{cor:compactness}
Suppose that $\Gamma$ is connected and that all endomorphisms
of $\Gamma$ are automorphisms of $\Gamma$.
Then either the relation $\Dist_{\{k\}}$ is pp-definable in $\Gamma$ for every $k \geq 1$, or the relation $\Diff_{\{k\}}$ is pp-definable in $\Gamma$ for every $k \geq 1$.
\end{Corollary}

\begin{Proof}
First consider the case that $\Gamma$ is preserved by the unary operation $x \mapsto -x$, and let $k \geq 1$ be arbitrary. Let $a_1,a_2$ be any
two elements of ${\mathbb Z}$ at distance $k$.
Since all endomorphisms of $\Gamma$ are
automorphisms of $\Gamma$, they are automorphisms of 
$({\mathbb Z}; \symsuc)$ by the
first statement of Theorem~\ref{thm:endos}. Hence we may apply Proposition~\ref{prop:compactness}, and there is a finite
set $S \subseteq {\mathbb Z}$ such that every homomorphism $f$
from $\Gamma[S]$ to $\Gamma$ satisfies $d(f(a_1),f(a_2))=d(a_1,a_2)$. Let $\phi(a_1,a_2)$ be the primitive positive formula
obtained from the canonical query for $\Gamma[S]$ by existentially
quantifying all vertices except for $a_1$ and $a_2$. We claim that
$\phi$ is a pp-definition of $\Dist_{\{k\}}$.

The relation
defined by $\phi$ contains the pair $(a_1,a_2)$ (since 
the identity mapping is a satisfying assignment for the canonical query
$\Gamma[S]$), and since $\Gamma$ is preserved
by all automorphisms of $({\mathbb Z}; \symsuc)$ it
also contains all other pairs $(x,y) \in {\mathbb Z}^2$ such that 
$d(x,y)=k=d(a_1,a_2)$. Conversely, $\phi$ does not contain 
any pair $(x,y)$ with $d(x,y) \neq k$. Otherwise, there must be a assignment $f \colon S \rightarrow \mathbb Z$ that satisfies the canonical query and maps $a_1$ to $x$ and $a_2$ to $y$.
This assignment is a homomorphism, and therefore contradicts the assumption that $d(f(a_1),f(a_2))=d(a_1,a_2)$. This proves 
the claim.

Now consider the case that $\Gamma$ is \emph{not} preserved
by the unary operation $-$. Again we use Theorem~\ref{thm:endos} and this time Proposition~\ref{prop:compactness2} to construct a primitive positive formula $\phi$
that defines the relation 
$\Diff_{\{k\}}$. 
\end{Proof}

\begin{Proposition}\label{prop:hard}
Suppose that for all $k$  the relation $\Dist_{\{k\}}$ 
is pp-definable in $\Gamma$.
Then $\Csp(\Gamma)$ is NP-hard.
\end{Proposition}
\begin{Proof}
Observe that the primitive positive formula $\exists y \, (d(x,y)=1 \wedge d(y,z)=5)$ defines the relation  $\Dist_{\{4,6\}}$. 
The structure $({\mathbb Z}; \Dist_{\{4,6\}})$ 
decomposes into two copies of the structure $({\mathbb Z}; \Dist_{\{2,3\}})$. 
This structure has the endomorphism $x \mapsto x \mod 5$,
and the image induced by this endomorphism is a cycle of length 5, which has a hard CSP 
(this is well-known; for a much stronger result on undirected graphs, see Hell and Ne\v{s}et\v{r}il~\cite{HellNesetril}).
\end{Proof}

\vspace{1cm}

\begin{Proof} (of Theorem~\ref{thm:def-succ})
By Lemma~\ref{lem:disconnected}, we can assume 
without loss of generality that $\Gamma$ is connected.    
If $\Gamma$ does not have a finite range endomorphism, then by Theorem~\ref{thm:endos} there is
an endomorphism of $\Gamma$ whose range induces in $\Gamma$ a substructure $\Delta$ which is first-order definable in $({\mathbb Z}; \suc)$, and where all endomorphisms
are automorphisms. 
Being the homomorphic image of the connected structure
$\Gamma$, $\Delta$ must also be connected.
We now apply Corollary~\ref{cor:compactness} to $\Delta$.
If the relation $\Dist_{\{k\}}$ is pp-definable in $\Delta$ for every $k \geq 1$, then $\Csp(\Gamma)$ (which is equal to $\Csp(\Delta)$ since $\Gamma$ and $\Delta$ are homomorphically equivalent) is NP-hard by Proposition~\ref{prop:hard}.
Otherwise, by Corollary~\ref{cor:compactness}, 
the relation $\Diff_{\{k\}}$ and in particular the relation $\suc$ is pp-definable in $\Delta$.
\end{Proof}

\section{Tractability of Modular Max}\label{sect:modular}
This section discusses the distance CSPs that can
be solved in polynomial time. 

	\begin{Definition}
	For $d \geq 1$, the \emph{$d$-modular max} is the operation $\max_d\colon\Z^2\to\Z$
	that is defined by $\max_d(x,y)=\max(x,y)$ if $x=y\mod d$ and $\max_d(x,y)=x$ otherwise.
	The \emph{$d$-modular min} is similarly defined as the operation $\min_d\colon\Z^2\to\Z$
	which satisfies $\min_d(x,y)=\min(x,y)$ if $x=y\mod d$ and $\min_d(x,y)=x$ otherwise.
	\end{Definition}
	
The results of this section 
improve the algorithmic
results that have been presented in the conference
version of the present paper~\cite{BodDalMarPin}. 

	\begin{Theorem}
	\label{thm:tractability-modular-semilattice}
	Suppose that $\Gamma$ has a $d$-modular max or $d$-modular min polymorphism.
	Then $\Csp(\Gamma)$ is in P.
	\end{Theorem} 
	The proof
of Theorem~\ref{thm:tractability-modular-semilattice}
can be found at the end of the section. 
	The first algorithm \textsc{Solve-Semilattice} we present in this section solves the CSP of structures
	that are closed under a \emph{semilattice operation}. 
	A semilattice operation
	is a binary function $f\colon\Z^2\to\Z$ that is idempotent, commutative, and associative.
	Note that for $d=1$, 
	the $d$-modular $\max$ equals the maximum operation, which is a semilattice operation. The situation for the $d$-modular $\min$ is dual, and we therefore restrict our discussion in this section to the $d$-modular $\max$ in the following.  
	
		Also note that the $d$-modular $\max$ is not commutative when $d>1$, \mbox{i.e.}, $\max_d$ is a semilattice operation only when $d=1$.
	We first treat the special case $d=1$, i.e.,
	the case that $\Gamma$ is preserved by
	$\max$, 
	using the technique of \emph{sampling} presented in~\cite{BodMacpheTha}.
	This case will then be used later to solve the general case. 
	
	\begin{Definition}
		Let $\Delta$ be a relational structure. 
		A \emph{sampling algorithm} 
		for $\Delta$
		is an algorithm that takes as input a natural number $n$ and returns a finite induced
		substructure $\Sigma$ of
		$\Delta$ such that for all instances $\Phi$ of $\Csp(\Delta)$
		with at most $n$ variables,
		we have $\Delta \models \Phi$ if and only if $\Sigma\models\Phi$.
	\end{Definition}

	\begin{Theorem}[Theorem 2.4 from~\cite{BodMacpheTha}]
		Let $\Delta$ be a structure over a finite relational signature with a semilattice polymorphism.
		If there exists a polynomial-time sampling algorithm for $\Delta$, 
		then $\Csp(\Delta)$ is in P.
	\end{Theorem}

	Thus, in order to obtain the polynomial-time
	algorithm \textsc{Solve-Semilattice} for $\Csp(\Gamma)$ when $\Gamma$ is preserved by $\max$, 
	it remains to prove that we can efficiently sample from $\Gamma$.
	Note that $\phi(x_1,\dots,x_n)$ is satisfiable in $\Gamma$ iff it is satisfiable in the substructure 
	induced by $\Gamma$ on $\{0,\dots,(D+1)n\}$.
	The sampling algorithm for $\Gamma$ then simply returns this structure,
	and the running time is polynomial in $n$.

	We now present 
	a more general algorithm that solves the CSP
	of any structure fo-definable in $(\Z;\suc)$ that is preserved by
	at least one of $d$-modular max. We need the following concept, which is important also in Section~\ref{sect:classification}. 


\begin{Definition}\label{def:progression}
A set of the form 
$[a,b]_d := \{a,a+d,a+2d,\ldots,b\}$
will be called an \emph{arithmetic $d$-progression}. 
A \emph{$d$-progression} is a binary relation of the form 
$\mathrm{Diff}_{[a,b]_d}$ for $a \leq b$ and $b - a$ divisible by $d$.
\end{Definition}

A binary relation is called \emph{trivial} if it is pp-definable in $(\Z;\suc)$, and \emph{non-trivial} otherwise. 
Note that $\mathrm{Diff}_{[a,b]_d}$
is non-trivial if and only if $a<b$.  
An arithmetic $d$-progression
$[a,b]_d$ is called \emph{non-trivial} if
$a<b$. 

\begin{Lemma}\label{lem:d-progression}
Suppose that $\Gamma$ is preserved
by $\max_d$, and that $R$ is
a non-trivial binary relation that is pp-definable in $\Gamma$. 
Then $R$ is a $d$-progression. 
\end{Lemma}
\begin{Proof}
Suppose for contradiction that $R$ is not a $d$-progression. There are two cases. 
\begin{itemize}
\item $R$ contains $(0,a)$ and $(0,b)$ where $a \neq b \mod d$. 
Let $\phi(x,y)$ be the primitive positive 
definition of $R$ in $\Gamma$.
As $R$ is distinct from 
${\mathbb Z}^2$, in the canonical database of $\phi$ the vertices $x$ and
$y$ must lie in the same connected
component, and it follows that $R$ has finite degree. 
Therefore, and because $R$ is non-trivial, there exists a smallest $p \in {\mathbb Z}$ so that $(0,p) \in R$. Choose some
$p'>p$ so that $(0,p') \in R$ and $p \neq p' \mod d$, which exists by assumption. Applying $\max_d$ to $(0,p)$ and $(d,d+p')$ we get
$(d,p) \in R$ and therefore $(0,p-d) \in R$, in contradiction to the choice of $p$. 
\item Suppose that case 1 does not apply. Then $R$ must contain some $(0,a)$ and $(0,c)$ but not $(0,b)$ for $0<a<b<c$ (with $a=b=c \mod d$).  
In this case, choose $a,b,c$ so that $c$ is minimal. 
Applying $\max_d$ to $(0,c)$ and $(d,d+a)$ we get $(d,c) \in R$ in contradiction to the minimality of $c$. 
\end{itemize}
In both cases we reached a contradiction, so $R$ must indeed be a $d$-progression. 
\end{Proof}

	\begin{Lemma}
	\label{lem:reduction-all-equal-mod-d}
	Let $d$ be a positive integer and let $\Gamma$ be such that the non-trivial binary relations pp-definable
	in $\Gamma$ are $d$-progressions. 
	Then there is an fo-expansion $\Delta$ of $(\Z;\suc)$ such that
	\begin{itemize}
	\item every relation of $\Delta$ is pp-definable in $\Gamma$;
	\item every relation of $\Gamma$ is pp-definable in $\Delta$;
	\item each relation $R$ of $\Delta$ but $\suc$ satisfies that for all $(a_1,\dots,a_n)\in R$,
	we have $a_1=a_i\mod d$ for all $1\leq i\leq n$.
	\end{itemize}
	\end{Lemma}
	\begin{Proof}
	Let $R$ be a relation of $\Gamma$, of arity $n$.
	It follows from the fact that the non-trivial binary relations pp-definable in $\Gamma$ are $d$-progressions
	that for all $(a_1,\dots,a_n),(b_1,\dots,b_n)\in R$, we have that $a_i-a_1=b_i-b_1\mod d$ for each 
	$i \in \{1,\dots,n\}$, and let $p_i\in \{0,\dots,d-1\}$ be this quantity. Define $R'(y_1,\dots,y_n)$ by
	$\exists x_2,\dots,x_n \, \big (R(y_1,x_2,x_3,\dots,x_n) \land \bigwedge_{i\geq 2} x_i = s^{p_i}(y_i) \big )$. 
	Then it is easy to check that if $(c_1,\dots,c_n)$ is a tuple in $R'$,
	then we have $c_i=c_1\mod d$ for all $1\leq i\leq n$.
	Let $\Delta = (\Z; R'_1,\dots,R'_m, \suc)$, where $R'_j$ is the relation constructed as above
	from $R_j\in\Gamma$.
	The claim about the form of the relations of $\Delta$ easily follows from our construction.
	\end{Proof}

\vspace{.4cm}
	We say that 
	$\Gamma$ is 
	\emph{$d$-nice} 
	if it satisfies the third item 
	in Lemma~\ref{lem:reduction-all-equal-mod-d}
	for some positive integer $d$. 
	Per the lemma, it suffices to do the complexity classification for $d$-nice
	structures $\Gamma$. 

	\begin{Lemma}
	\label{lem:modular-max-max}
	Let $\Gamma$ be $d$-nice.
	$\Gamma$ is preserved by $\max_d$ 
	if and only if $\Gamma/d$ is preserved by $\max$. 
	\end{Lemma}
	\begin{Proof}
	Let $R$ be a relation of $\Gamma$ that is not $\suc$ ($\suc$ is preserved by the four
	operations mentioned in the statement so there is nothing to prove for this relation).

	(Forwards.) Let $(a_1,\dots,a_n),(b_1,\dots,b_n)$ be tuples in $R$ such that $a_i=b_i=0\mod d$ for all $i$.
	Then we have that $(\max(a_1,b_1),\dots,\max(a_n,b_n)) = (\max_d(a_1,b_1),\dots,\max_d(a_n,b_n))$, and this tuple is in $R$ by hypothesis.
	Furthermore, the entries of this tuples are divisible by $d$, and thus the tuple belongs to $R^{\Gamma/d}$.

	(Backwards.) Let $(a_1,\dots,a_n),(b_1,\dots,b_n)$ be tuples in $R$. Note that $a_i-a_j=b_i-b_j=0\mod d$ for all $i,j$.
	Thus, if $a_i\neq b_i\mod d$ for some $i$, then $a_i\neq b_i\mod d$ for \emph{all} $i$
	and in this case $(\max_d(a_1,b_1),\dots,\max_d(a_n,b_n))=(a_1,\dots,a_n)$, which is in $R$.
	Otherwise, $a_i=b_i\mod d$ for all $i$, and thus the two tuples $(0,a_2-a_1,\dots,a_n-a_1)$ and $(b_1-a_1,\dots,b_n-a_1)$
	are in $R$ and have all their entries divisible by $d$.
	Hence, $(\max_d(0,b_1-a_1),\dots,\max_d(a_n-a_1,b_n-a_1))$ is in $R^{\Gamma/d}$, since $\Gamma/d$ is preserved by $\max$.
	It follows that $(\max_d(0,b_1-a_1),\dots,\max_d(a_n-a_1,b_n-a_1))+a_1 = (\max_d(a_1,b_1),\dots,\max_d(a_n,b_n))$ is in $R$,
	and $R$ is preserved by $\max_d$.
	\end{Proof}

\vspace{.4cm}
	Suppose $\Gamma$ is $d$-nice
	and has the relations $R_1,\dots,R_m$
	and $\suc$, 
	and let $\Phi = \exists x_1,\dots,x_n.\phi$
	be an instance of $\Csp(\Gamma)$.
	Our algorithm works as follows:
	\begin{enumerate}
		\item Compute the finest equivalence relation on the set of variables $V$ with parts
		$V_1,\dots,V_\ell$ so that
		there is no constraint $\suc(x,y)$ in $\phi$ when $x,y$ are in the same subset,
		and so that if a constraint $R_i(y_1,\dots,y_k)$ is in $\phi$
		then $y_1,\dots,y_k$ are all in the same equivalence class, and if $y_1=y_2$ is in $\phi$ then $y_1$ and $y_2$ are in the same equivalence class. If no such partition exists, reject $\Phi$. It is clear that this computation can be performed in polynomial time in the size of the input. 
		\item Build a primitive positive sentence $\Psi$ that contains an existentially
		quantified variable $v_i$ for each subset $V_i$,
		and that contains a 
		conjunct $\suc(v_j,v_i)$ iff
		there exist variables $x\in V_i,y\in V_j$ so that $\suc(y,x)$
		is a conjunct in $\phi$.
		\item Test whether $\Psi$ is true in 
		$\vec{C}_d$,
		the directed cycle with the vertex set $\{0,1,\dots,d-1\}$ (we also use $\suc$ to denote the edge predicate in this structure).
		Reject $\Phi$ if $\Psi$ is not true in 
		$\vec{C}_d$.
		\item Otherwise let $p \colon\{v_1,\dots,v_\ell\}\to \vec{C}_d$
		be a satisfying assignment to the quantifier-free part of $\Psi$. 
		\item Define a new sentence $\Xi$ as follows: for each variable $x$ in $V_i$ add
		an (existentially quantified) variable $z_{x}^{-p(i)}$.
		For each constraint $R(y_1,\dots,y_k)$ 
		 with $y_1,\dots,y_k\in V_i$ and $R \in \{R_1,\dots,R_m\}$ 
		we add the constraint $R(z_{y_1}^{-p(i)},\dots$,
		$z_{y_k}^{-p(i)})$.
		Finally, for each constraint $\suc(y,x)$ with $x\in V_j$, $y\in V_i$, if $p(j)>0$ 
		add the constraint $z_{x}^{-p(j)}=z_{y}^{-p(i)}$ and if $p(j) = 0$ (which means that $p(i) = d-1$) 
		add $\suc$-constraints to express that $z_{x}^{0} = s^d(z_y^{-p(i)})$.
		\item Run \textsc{Solve-Semilattice} on $\Xi$, as an input to $\Csp(\Gamma/d)$,
		and accept $\Phi$ iff \textsc{Solve-Semilattice} accepts $\Xi$.
	\end{enumerate}
	
	Note that at the steps $3$ and $4$, we need a polynomial-time algorithm
	that solves $\Csp(\vec{C}_d)$ and that also builds a solution.
	It is well-known that a greedy approach works here,
	which we describe below for the sake of
	completeness. Assign the first variable $x$ 
	to any vertex of $\vec{C}_d$. At each step, if the variables in $V$ have already been assigned,
	consider a variable $y$ 
	such that there exist $x\in V$ 
	and an atomic formula $\suc(x,y)$.
	If all such variables $x$ are assigned to the same value, then assign $y$ to the next vertex in $\vec{C}_d$.
	Otherwise, reject the instance.
	It is clear that this algorithm builds a satisfying assignment if and only if
	a satisfying assignment exists.

	\begin{Lemma}
	\label{lem:equisatisfiability-factor}
	The formula $\Xi$ is true in $\Gamma$
	if and only if it is true in $\Gamma/d$.
	\end{Lemma}
	\begin{Proof}
		If $\Xi$ is true in $\Gamma$,
		we may assume by translation that for all $i$ at least one of the 
		variables of $V_i$ is mapped to some integer in $\Gamma/d$.
		Since all the relations of $\Gamma$ 
		with 
		the exception of $\suc$
		only contain tuples $(a_1,\dots,a_r)$ with $a_i=a_j\mod d$ for all $1\leq i < j \leq r$,
		and since the $\suc$ constraints in $\Xi$
		only appear to express formulas
		of the form $y = s^d(x)$, 		
		it follows that all the variables are actually mapped to $\Gamma/d$.
		The converse is trivial, $\Gamma/d$ being an induced substructure of $\Gamma$.
	\end{Proof}
\vspace{.4cm}

We finally prove the main result of this section, 
Theorem~\ref{thm:tractability-modular-semilattice}. 
\vspace{.4cm}

	\begin{Proof}
	By Lemma~\ref{lem:d-progression}, all non-trivial binary relations with a primitive positive definition in $\Gamma$ are $d$-progressions. 
	By Lemma~\ref{lem:reduction-all-equal-mod-d}, we can assume without loss
	of generality that $\Gamma$ is $d$-nice. Let $\Phi = \exists x_1,\dots,x_n. \,\phi$ 
		be an instance of $\Csp(\Gamma)$,
		and suppose that $\Phi$ is true in $\Gamma$.
		Let $h \colon \{x_1,\dots,x_n\}\to\Z$ be a satisfying assignment of $\phi$.
		Then congruence of the $h$-image of a variable modulo $d$ 
		defines an equivalence relation on $V$ with the properties required in the first item of the algorithm, so the algorithm does not reject at this step. 
		
		We prove first that $\Psi$ is true in $\vec{C}_d$, and thus that $\Phi$ is not
		rejected by the algorithm at Step 3. If $v_1,\dots,v_k$ are the variables
		in $\Psi$, define $t(v_i) = h(x) \bmod d$ for any $x\in V_i$.
		This is well-defined, for if two variables $x,y$ are in the same set $V_i$,
		there are tuples of variables ${\mathbf a}^1,\dots,{\mathbf a}^m$
		with $x$ being an element of ${\mathbf a}^1$, $y$ 
		being an element of ${\mathbf a}^m$
		and with constraints $R({\mathbf a}^k)$ in $\phi$ for all $1\leq k\leq m$.
		Since $\Gamma$ is $d$-nice,
		it must be that $h(x)=h(y) \bmod d$, so that $t(v_i)$ is well defined.
		If there is a constraint $\suc(v_j,v_i)$ in $\Psi$,
		there is a corresponding constraint $\suc(y,x)$ with $x\in V_i,y\in V_j$.
		Thus, we have that $h(x)=h(y)+1$, which entails that $t(v_i)=t(v_j)+1\bmod d$,
		and $t$ satisfies the constraint in $\Phi$.
		Therefore, $\Psi$ is true in $\vec{C}_d$.
		
		We now prove that the sentence $\Xi$ (computed by the algorithm) is true in $\Gamma$ (and hence in $\Gamma/d$,
		by Lemma~\ref{lem:equisatisfiability-factor}), and thus that $\Phi$
		is accepted by the algorithm.
		Define $r(z_x^{-p(i)}) :=h(x)-p(i)$ (which explains the notation we employed).
		We claim that $r$ satisfies the constraints in $\Xi$.
		If $R(z_{y_1}^{-p(i)},\dots,z_{y_k}^{-p(i)})$
		for $R \in \{R_1,\dots,R_m\}$ is a constraint in $\Xi$,
		then $R(y_1,\dots,y_k)$ is a constraint in $\Phi$,
		so that we have $\Gamma\models R(h(y_1),\dots,h(y_k))$.
		As a consequence, we have $\Gamma\models R(h(y_1)-p(i),\dots,h(y_k)-p(i))$
		since translations preserve $\Gamma$.
		Noting that $h(y_l)-p(i) = r(z_{y_l}^{-p(i)})$, we have $\Gamma\models R(r(z_{y_1}^{-p(i)}),\dots,r(z_{y_k}^{-p(i)}))$.
		It remains to be checked that the equality constraints are satisfied by $r$.
		Let $z_{x}^{-p(j)}=z_{y}^{-p(i)}$ be such an equality constraint,
		and let $\suc(y,x)$ be the corresponding constraint in $\phi$, with $x \in V_j$, $y \in V_i$, and $p(j)>0$.
		By the properties of $p$, we have that $p(j)=p(i)+1$,
		and it follows from $h(x) = h(y)+1$ that $h(x)-p(j) = h(y)-p(i)$, 
		i.e., $r(z_x^{-p(j)})=r(z_y^{-p(i)})$.
		If $z_{x}^{0} = s^d(z_y^{-p(i)})$ is in $\Xi$,
		then $\suc(y,x)$ is in $\Phi$ with $y\in V_{d-1}$, $x\in V_j$, and $p(j) = 0$. 
		As a consequence, from $h(x)=h(y)+1$ follows that 
		$r(z_x^0) = r(z_{y}^{-(d-1)})+d-1+1 =  r(z_{y}^{-(d-1)})+d$.
		
		Let us now prove that if the algorithm accepts $\Phi$,
		then $\Phi$ is indeed true in $\Gamma$.
		Let $r$ be an assignment that satisfies the constraints in $\Xi$.
		For $x\in V_i$, define $h(x) := r(z_{x}^{-p(i)})+p(i)$.
		If $R(y_1,\dots,y_k)$ is a constraint in $\phi$
		with all the variables in $V_i$ and $R \in \{R_1,\dots,R_m\}$, then
		$R(z_{y_1}^{-p(i)},\dots,z_{y_k}^{-p(i)})$ is a constraint in $\Xi$
		so that $R(r(z_{y_1}^{-p(i)}),\dots,r(z_{y_k}^{-p(i)}))$ holds in $\Gamma$,
		and by translation we have that $R(r(z_{y_1}^{-p(i)})+p(i),\dots,r(z_{y_k}^{-p(i)})+p(i))$
		also holds in $\Gamma$.
		If $\suc(y,x)$ is a constraint in $\Phi$,
		then we have $x\in V_j$, $y\in V_i$, and $p(j)=p(i)+1 \bmod d$.
		If $p(j)=0$ then the constraint $z_{x}^{0}=s^d(z_{y}^{-d+1})$ is in $\Xi$,
		so that $r(z_{x}^{0}) = r(z_{y}^{-(d-1)})+(d-1)+1$, i.e., $h(x) = h(y)+1$.
		If $p(j)>0$, the constraint $z_{x}^{-p(j)}=z_{y}^{-p(i)}$ is in $\Xi$,
		so that $h(x)=r(z_{x}^{-p(j)})+p(j) = r(z_{y}^{-p(i)})+p(j) = h(y)-p(i)+p(j)$,
		and by Step 3, we have $p(j)=p(i)+1$ so that $h(x)=h(y)+1$.	
	\end{Proof}

\section{Classification}
\label{sect:classification}

In this section we finish the complexity classification for those
$\Gamma$ that do not have a finite core.
The main result of Section~\ref{sect:def-succ} 
shows that, unless $\Gamma$ has a finite core, for the complexity classification of $\Csp(\Gamma)$ we can assume that the structure $\Gamma$ contains
the relation $\suc$.
In the following we therefore assume that the structure $\Gamma$
contains the relation $\suc$; moreover, we freely use expressions
of the form $y-x=d$, for fixed $d$, in primitive positive definitions
since such expressions have themselves pp-definitions from $\suc$ 
and therefore from $\Gamma$. 
Our main result will be the following.

\begin{Theorem}\label{thm:class}
Suppose that $\Gamma$ contains the relation $\suc$.
Then $\Gamma$ is preserved by a modular max or modular min and $\Csp(\Gamma)$ is in P, or $\Csp(\Gamma)$ is NP-hard.
\end{Theorem}

An $n$-ary relation $R$ on a set $X$ is \emph{$r$-decomposable} if 
$R$ contains all $n$-tuples $(a_1,\dots,a_n)$
such that  for every $r$-element subset $I$ of $\{1,\dots,n\}$
there is a tuple $(b_1,\dots,b_n) \in R$ such that $a_i = b_i$ for all $i \in I$.


\begin{Lemma}
\label{lem:positive-not-modular-binary-relation}
	Suppose that $\Gamma$ contains
	the relation $\suc$ and 
	does not admit a modular max or modular min polymorphism.
	Then there is a pp definition in $\Gamma$ of a non-trivial binary relation of finite degree.
\end{Lemma}
\begin{Proof}
 Assume for contradiction that the binary 
 relations pp-definable in $\Gamma$
	are already pp-definable in $(\Z;\suc)$. 
	If every relation $S$ pp-definable in $\Gamma$ were $2$-decomposable, 
	then $S$ would be invariant under a modular max or modular min operation, since the $2$-decomposable
	relations that have a pp-definition in $\Gamma$ already have a pp-definition
	in $(\Z;\suc)$, which means that they are
	preserved by the $d$-modular max and  $d$-modular min for all $d\geq 1$.
	Hence, there is a relation pp-definable in $\Gamma$ that is not $2$-decomposable.
	Let $R$ be such a relation of smallest possible 
	arity $r \geq 3$. 
	In particular, $R$ is not $(r-1)$-decomposable, 
	and hence there exists a tuple $(a_1,\ldots,a_r) \notin R$ such that for all $i \in [r]$, $(a_1,\ldots,p_i,\ldots,a_r) \in R$
	for some integer $p_i$.
	By replacing $R$ by the pp-definable relation
	\[ \exists y_1,\ldots, y_r \, \big(R(y_1,\ldots, y_r) \wedge \bigwedge_{i \in [r]} (y_i = x_i + a_i) \big )  \]
	we can further assume that $a_i = 0$ for all $i \in [r]$.
	We can also assume, w.l.o.g., that $p_1\neq -p_2$ because $r\geq 3$.

	Suppose that the arity of $R$ is greater than $3$,
	and consider now the ternary relation $T(x_1,x_2,x_3)$
	defined by $R(x_1,x_2,x_3,\ldots,x_3)$.
	Suppose there is a $z$ so that $R(0,0,z,\ldots,z)$,
	then $T$ would not be $2$-decomposable since $(0,0,0)\not\in T$, although
	$(p_1,0,0),(0,p_2,0)$, and $(0,0,z)$ are all in $T$, which contradicts the
	minimality of the arity of $R$.
	If there is no such $z$ then
	$\exists x_3.R(x_1,x_2,x_3,\ldots,x_3)$
	defines a binary relation omitting $(0,0)$ and containing $(0,-p_1)$ and $(0,p_2)$.
	This relation is non-trivial, a  contradiction.

	Thus we are in the situation in which $r=3$.
	If a binary projection of $R$ is non-trivial, 
	we are done, so suppose that all binary
	projections are trivial. We claim that 
	every binary projection of $R$ must in fact be $\Z^2$:
	otherwise one such binary projection, w.l.o.g.\ $\exists x_1. R(x_1,x_2,x_3)$, 
	would be equivalent to $x_3=x_2+p$ for some $p\in\Z$.
	Let $(a,b,c)$ be such that $(a,b)$ is in the projection of $R$ along coordinates $\{1,2\}$,
	$(a,c)$ is in the projection of $R$ along $\{1,3\}$,
	and $(b,c)$ is in the projection of $R$ along $\{2,3\}$ (i.e., $c=b+p$).
	Since $(a,b)$ is in the first projection of $R$, there exists $d\in\Z$
	such that $(a,b,d)$ is in $R$, but since the third projection is trivial
	we have $d=b+p=c$, so that $(a,b,c)$ is in $R$ and $R$ is $2$-decomposable,
	contradicting our assumptions. Thus every binary projection of $R$ is $\Z^2$.
	
	A formula over the signature of $(\Z; s)$ 
	in disjunctive normal form (DNF) is called \emph{reduced} when every formula obtained by removing literals or clauses is not logically equivalent over $(\Z; s)$, and 
	if every atomic formula is of the form
	$y = s^n(x)$ for $n \in \mathbb N$. 
	Let $\phi(x_1,x_2,x_3)$ be a formula in reduced DNF that defines $R$.
	This formula has at least two disjuncts, otherwise $R$ would be pp-definable
	over $(\Z;s)$.
	We claim that there is a disjunct in $\phi$ that consists of only one literal.
	If that was not the case, every disjunct ${\mathcal D}_i$ would be equivalent
	to $x_1=s^{p_i}(x_2)\land x_1=s^{q_i}(x_3)$ for some $p_i,q_i\in\Z$ (for negative $p$, the expression $x = s^p(y)$ is notational sugar for $y = s^{-p}(x)$).
	In this case, the formula $\exists x_2. \phi(x_1,x_2,x_3)$ defines a binary non-trivial finite-degree relation, contradicting what we proved in the previous paragraph.
	Furthermore, there are at least two such disjuncts:
	if there is only one, say $x_1=s^p(x_2)$, then the relation defined
	by $\exists x_3. \phi(x_1,x_2,x_3)$ is binary non-trivial finite-degree, a contradiction.
	Hence there are two disjuncts in $\phi$, which are up to renaming the variables
	$x_1=s^p(x_2)$ and $x_1=s^q(x_3)$.
	Then the formula $\exists x_3 \, \big(\phi(x_1,x_2,x_3) \land x_3=s^{p-q+1}(x_2)\big)$
	is equivalent to a formula in DNF which is reduced and contains the two disjuncts
	$x_1=s^p(x_2)$ and $x_1=s^{p+1}(x_2)$.
	The relation defined by this formula proves the lemma.
\end{Proof}

\begin{Proposition}
Let $a, b$ be two odd numbers such that $a<b$. Then the problem $\Csp({\mathbb Z}; \suc, \Diff_{\{0,a,b,a+b\}})$ is NP-hard.
\end{Proposition}
\begin{Proof} 
Let $k$ be the integer $\frac{a+b}{2}$. 
Note that the pp-formula 
$$\phi(x,z) = \exists y \, (\Diff_{\{0,a,b,a+b\}}(x,y) \wedge y-z=k)$$ defines the relation $C :=\Dist_{\{\frac{b-a}{2},\frac{b+a}{2}\}}= \big \{(x,z) \; | \; d(x,z) \in \{\frac{b-a}{2},\frac{b+a}{2}\} \big \}$. 
Consider the mapping $f \colon {\mathbb Z}\rightarrow \{0,\dots,b-1\}$ defined by $f(x)=x \mod b$. It follows from $\frac{b-a}{2}=-\frac{b+a}{2} \mod b$ that $f$ preserves $C$. It also follows by the same reason that the restriction of $C$ to $\{0,\dots,b-1\}$ defines a graph $D$ where every node has two edges. 
Furthermore, if $m$ is $\gcd(\frac{b-a}{2},\frac{b+a}{2})$ then $D$ is the disjoint union of $m$ cycles of $\frac{b}{m}$ nodes. Since $\frac{b}{m}$ is odd we
have that $\Csp({\mathbb Z}; C)$ is NP-hard (this follows from~\cite{HellNesetril}).
\end{Proof}

\begin{Lemma}
\label{lem:basichardcase}

Let $a,b,c \in \mathbb Z$ with $b \neq c$. Then $\Csp({\mathbb Z}; \suc, \Diff_{\{a,b\}}, \Diff_{\{a,c\}})$ is NP-hard.
\end{Lemma}
\begin{Proof}
First observe that the pp-formula $\exists u \, (\Diff_{\{a,b\}}(x,u) \wedge u=y+a)$
defines the relation $\Diff_{\{0,b-a\}}$; similarly, there is a pp-definition
of $\Diff_{\{0,c-a\}}$ in $({\mathbb Z}; \suc, \Diff_{\{a,b\}}, \Diff_{\{a,c\}})$. 
Let $d=b-a$ and $e=c-a$; we will show that
$\Csp({\mathbb Z}; \suc, \Diff_{\{0,d\}}, \Diff_{\{0,e\}})$ is NP-hard.

The relation defined by $\exists u \, (\Diff_{\{0,d\}}(x,u) \wedge \Diff_{\{0,e\}}(u,y))$ is $\Diff_{\{0,d,e,d+e\}}$. 
If both $d$ and $e$ are odd, 
we obtain hardness of the CSP 
from the previous proposition applied to $({\mathbb Z}; \suc, \Diff_{\{0,d,e,d+e\}})$. If both $d$ and $e$ are even, 
then the structure $\Delta := ({\mathbb Z}; \Diff_{\{0,d\}},\Diff_{\{0,e\}},\{(x,y) \; | \; x-y=2\})$ is pp-definable in $\Gamma$. 
The structure $\Delta$ is isomorphic to the disjoint union of two
copies of the structure $({\mathbb Z}; \suc,\Diff_{\{0,d/2\}},\Diff_{\{0,e/2\}})$;
the claim now follows by induction on the minimum even among $d$ and $e$ (the case where they are both odd being already solved).

Finally, assume that precisely one of $d$ or $e$ is even; say $d$ is even. Write $\lcm(d,e)$ for the least common
multiple of $d$ and $e$, and set $u:=\lcm(d,e)/d$ and
$v:=\lcm(d,e)/e$. The formula 
\begin{align*} \exists y_1, \dots, y_u, z_1, \dots, z_v \big ( & \Diff_{\{0,d\}}(p,y_1) \wedge \Diff_{\{0,d\}}(y_1,y_2) \wedge \dots \wedge 
\Diff_{\{0,d\}}(y_{u-1}, q) \\
\wedge & \Diff_{\{0,e\}}(p,z_1) \wedge \Diff_{\{0,e\}}(z_1,z_2) \wedge \dots \wedge 
\Diff_{\{0,e\}}(z_{v-1},q) \big )
\end{align*}
with free variables $p$ and $q$ defines $\Diff_{\{0,\lcm(d,e)\}}$.
We are now again in the case 
that we can pp-define two relations $\Diff_{\{0,g\}}$ and $\Diff_{\{0,h\}}$ for even $g,h$
(namely, $g=d$ and $h=\lcm(d,e)$), and thus we are done.
\end{Proof}

\begin{Lemma}
\label{lem:largeprogression}
Let $S$ be a finite set of integers with $|S|>1$ with elements of the form  $i\cdot d$ where $i\in{\mathbb Z}$. 
Let $md=\min(S)$, $Md=\max(S)$, let $[jd,kd]_d \subseteq S$ be maximal, let $l$ be such that $l\geq \max(j-m-1,M-k-1,0)$
and such that $k\geq j+l$. Then every $d$-progression with at most 
$r := k-j-l+1$ elements is pp-definable in $({\mathbb Z}; \suc, \Diff_S)$. 
\end{Lemma}
\begin{Proof}
We shall show first how to pp-define a $d$-progression $\Diff_T$ where $T \subseteq {\mathbb Z}$ has {\em exactly} $r=k-j-l+1$ elements. 
For every $0\leq i\leq l$, let $\phi_i(x,y)$ be the formula 
$\exists z \, (z=x+id
\wedge \Diff_S(z,y))$ which is equivalent to a pp-formula over the relations $\Diff_S$ and $\suc$. There exists $T \subseteq \Z$  such that 
$\Diff_T$ is the relation defined by 
$\phi := \bigwedge_{0\leq i\leq l} \phi_i(x,y)$. 
We claim that $T$ is precisely $[(j+l)d,kd]_d$. 

We have $T \subseteq S$ because the formula contains the conjunct $\phi_0(x,y)$. Let $s=nd$ be any element of $S$. Let us do a case analysis.

\begin{enumerate}
\item Case $m \leq n<j-1$. 
In this case $0 \leq j-m-1 \leq l$, and thus $\phi$
contains the conjunct
$\phi_{j-m-1}(x,y) = \exists z \, \big(z = x + (j-m-1) d \wedge \Diff_S(z,y) \big)$. 
The smallest $y$ such that $\phi_{j-m-1}(0,y)$ 
holds is $(j-1)d$. Hence, in this case $s \not\in T$.
\item Case $j-1\leq n<j+l$. By the maximality of $[jd,kd]_d$ it follows that $(j-1)d \not\in S$. Then 
$\phi_i(0,nd)$ does not hold if we pick $i=n-j+1$. Hence, $s\not\in T$.
\item Case $j+l\leq n\leq k$. For every $1\leq i\leq l$, we have that $\phi_i(0,nd)$ holds as $j+i\leq n\leq k+i$. This implies that $s \in T$.
\item Case $k<n\leq M$. By the maximality of $[jd,kd]_d$ we have that $(k+1)d\not\in S$. Hence, by choosing $i=n-(k+1)$ we have that
$\phi_i(0,s)$ does not hold. Consequently $s\not\in T$. 
\end{enumerate}
Hence, $T$ has exactly $r$ elements with largest element $kd$,
and $\Diff_T$ is a $d$-progression with the pp-definition $\phi$. Now, if 
$\Diff_P$ is a $d$-progression where
$P$ has exactly $r$ elements, then 
$\Diff_P$ can be defined by the
pp-formula $\exists z \, (z=x+p \wedge \phi(z,y))$
choosing $p:=\max(P)-kd$. 

Finally, we turn our attention to 
arithmetic $d$-progressions $T$
with less then $r$ elements. 
If $T$ has $r-1$ elements we can use the pp-formula $\exists z \, (z=x+d \wedge \phi(x,y) \wedge \phi(z,y))$
and apply some shift by successor 
to pp-define $\Diff_T$. Iterating the previous construction we can pp-define every $d$-progression. 
\end{Proof}

\vspace{.2cm}

Let us illustrate the construction of $\Diff_P$ 
in the previous proof with an example. Assume $S$ is the set 
$\{1,4,5,7,8,9,10,11,12,13,14,15,16,18,19,20\}$ 
which we can represent as:

{\tiny
\[\mathclap{%
\begin{array}{c| c | c | c | c | c | c | c | c | c | c | c | c | c | c | c | c | c | c | c | c | c | c | c | c | c | c |}%
               &1    &2  &3  &4  &  5& 6 & 7 & 8 & 9 & 10&11 &12 &13 &14 &15 &16 &17 &18 &19 & 20&21 &22 &23 &24 &25 &26  \\%
               \hline%
            S & \bu &   &   &\bu&\bu&   &\bu&\bu&\bu&\bu&\bu&\bu&\bu&\bu&\bu&\bu&   &\bu&\bu&\bu&   &   &   &   &   &      \\%
\end{array}}
\]
}
We have $d=1$. Consider $[7,16]_1 \subseteq S$. Then we have $m=1$, $M=20$, $j=7$, and $k=16$. Fix $l=6$. Then $r = 16 - 7 - 6 + 1 = 4$. 
For every $i \in \{0,\dots,l\}$ let $Z_i$ be such that $\phi_i$ defines $\Diff_{Z_i}$. The situation can be illustrated as follows.

{\tiny
\[\mathclap{%
\begin{array}{c|c|c|c| c | c | c | c | c | c | c | c | c | c | c | c | c | c | c | c | c | c | c | c | c | c | c |}%
               &1    &2  &3  &4  &  5& 6 & 7 & 8 & 9 & 10&11 &12 &13 &14 &15 &16 &17 &18 &19 & 20&21 &22 &23 &24 &25 &26    \\%
               \hline%
            Z_0& \bu &   &   &\bu&\bu&   &\bu&\bu&\bu&\bu&\bu&\bu&\bu&\bu&\bu&\bu&   &\bu&\bu&\bu&   &   &   &   &   &       \\%
            Z_1&     &\bu&   &   &\bu&\bu&   &\bu&\bu&\bu&\bu&\bu&\bu&\bu&\bu&\bu&\bu&   &\bu&\bu&\bu&   &   &   &   &       \\%
            Z_2&     &   &\bu&   &   &\bu&\bu&   &\bu&\bu&\bu&\bu&\bu&\bu&\bu&\bu&\bu&\bu&   &\bu&\bu&\bu&   &   &   &       \\%
            Z_3&     &   &   &\bu&   &   &\bu&\bu&   &\bu&\bu&\bu&\bu&\bu&\bu&\bu&\bu&\bu&\bu&   &\bu&\bu&\bu&   &   &      \\%
            Z_4&     &   &   &   &\bu&   &   &\bu&\bu&   &\bu&\bu&\bu&\bu&\bu&\bu&\bu&\bu&\bu&\bu&   &\bu&\bu&\bu&   &      \\%
            Z_5&     &   &   &   &   &\bu&   &   &\bu&\bu&   &\bu&\bu&\bu&\bu&\bu&\bu&\bu&\bu&\bu&\bu&   &\bu&\bu&\bu&      \\%
            Z_6&     &   &   &   &   &   &\bu&   &   &\bu&\bu&   &\bu&\bu&\bu&\bu&\bu&\bu&\bu&\bu&\bu&\bu&   &\bu&\bu&\bu    \\%
\hline%
            P  &     &   &   &   &   &   &   &   &   &\  &   &   &\bu&\bu&\bu&\bu&   &   &   &   &   &   &   &   &   &        \\%
\hline%
& \multicolumn{4}{c} {\text{ Case 1}} & & %
\multicolumn{6}{c} {\text{ Case 2}} & &%
\multicolumn{3}{c} {\text{ Case 3}} & &%
\multicolumn{3}{c} {\text{ Case 4}} & & & & & & &  \\%
\end{array}}
\]
}

\begin{Corollary}
\label{co:largeprogression}
Let $r>0$, let $S$ be a finite set of multiples of $d$, and assume that $[rd,3rd]_d \subseteq S$ and $S\subseteq[0,4rd]_d$. Then every $d$-progression $\Diff_T$ where $|T| \leq r$  is pp-definable in $({\mathbb Z}; \suc, \Diff_S)$.
\end{Corollary}
\begin{Proof}
Directly from Lemma~\ref{lem:largeprogression}. Note that the assumptions of the Corollary guarantee that $0\leq m$, $M\leq 4r$, $j\leq r$, and $3r\leq k$. It is straightforward to verify that $l=r$ gives the desired result.
\end{Proof}

\begin{Lemma}
\label{lem:generatedprogression}
Let $S$ be a finite set of integers with $|S|>1$ and let $d$ be the greatest common divisor of all $a-a'$ with $a,a'\in S$ and $a \neq a'$. 
Then any $d$-progression 
is pp-definable in $({\mathbb Z}; \suc, \Diff_S)$.
\end{Lemma}

\begin{Proof}
The set of 
non-trivial
maximal arithmetic $d$-progressions contained in $S$ can be totally ordered
by setting 
$T_1 \leq T_2$ if 
$\min(T_1) \leq \min(T_2)$. 
If $T_1<T_2$ then we define the distance from $T_1$ to $T_2$ to be $\min(T_2)-\max(T_1)$.

For any $m\geq 1$, let $(\Diff_S)^m$ be the relation $\overbrace{\Diff_S\circ \Diff_S\circ\cdots\circ \Diff_S}^m$ which we can
write as $\Diff_{S^m}$ where $S^m$ contains all integers that we can express as $a_1+\cdots+a_m$ with $a_1,\dots,a_m\in S$. Clearly, 
$(\Diff_S)^m$ is pp-definable from $\Diff_S$. 
By the definition of $d$ it follows that if $m$ is large enough there exists some integer $a$, such that $\{a,a+d\}\subseteq S^m$, or, in 
other words, that $S^m$ contains a non-trivial arithmetic $d$-progression. 
For ease of notation we shall assume that already $S$ contains 
a non-trivial arithmetic $d$-progression 
(otherwise replace $S$ by $S^m$).

Let $n$ be the maximum distance between
two consecutive arithmetic $d$-progressions contained
in $S$, and set $n=0$ if there is only one maximal arithmetic $d$-progression. 
Let $l^-$ (respectively $l^+$)
be minimal (respectively maximal) with the property that $\{\min(S^m)+l^-,\min(S)+l^-+d\}\subseteq S$ (respectively $\{\max(S)-l^+-d,\max(S)-l^+\}\subseteq S$). Finally,
define $l$ to be $\max(l^-,l^+)$.
Let $n_2$ and $l_2$ be defined as $n$ and $l$,
but with respect to $S^2$ instead of $S$.


{\em Claim 1}. $l_2\leq l$. 
\vspace{.1cm}

Proof: Follows from the fact that $\{2\min(S)+l^-,2\min(S)+l^-+d,2\max(S)-l^+-d,2\max(S)-l^+\}\subseteq S^2$.

\vspace{.1cm}
{\em Claim 2}. If $l=0$ then $n_2\leq n$. Furthermore, if $n>0$ the inequality is strict. 
\vspace{.1cm}

Proof: If $n=0$ then $S^2$ is necessarily an arithmetic $d$-progression and the claim follows. If $n>0$ then let $X<Y$ be consecutive non-trivial maximal arithmetic $d$-progressions contained in $S^2$ (such $X$ and $Y$ always exists if $n_2>0$, otherwise there is nothing to prove). We claim that there exist non-trivial maximal arithmetic $d$-progressions $A\leq B$ in $S$
such that $\max(A)+\max(B)\leq \max(X)$. Indeed, set $A=B$ to be the maximal arithmetic $d$-progression containing $\{\min(S),\min(S)+d\}$.
Consequently, we can choose $A\leq B$ satisfying the conditions of the claim with $\max(A)+\max(B)$ maximal. 

Since $X<Y$ it follows that $\max(A)<\max(S)$ which implies that there exists a 
non-trivial
maximal arithmetic $d$-progression $C$ 
in $S$ with $A<C$ (in particular consider the one containing $\{\max(S)-d,\max(S)\}$). Pick any such $C$ with $\min(C)$
minimal.

Since $S$ contains arithmetic $d$-progressions $A$ and $B$ it follows that $S^2$ contains the (not necessarily maximal) non-trivial arithmetic $d$-progression $[\min(A)+\min(B),\max(A)+\max(B)]_d$. Let $X'$ be a maximal arithmetic progression
in $S^2$ containing it. Similarly let $Y'$ be a maximal arithmetic $d$-progression in $S^2$ containing $[\min(B)+\min(C),\max(B)+\max(C)]_d$.
Since $\max(A)+\max(B)\leq \max(X)$ it follows that $X'\leq X$. Furthermore, by the maximality of $\max(A)+\max(B)$ and $A<C$ we have $X<Y'$. As $Y$ is consecutive to $X$ it follows that $Y\leq Y'$. The distance from
from $X'$ to $Y'$ is at most $\min(B)+\min(C)-\max(A)-\max(B)$, which, as we will show, is strictly smaller than $n$. Indeed, since $C$ is consecutive to $A$, we have
$\min(C)-\max(A)\leq n$ and since $B$ is non-trivial $\max(B)-\min(B)>0$.  This finishes the proof of Claim 2.

We can assume that $\min(S)=0$ by applying some shift by successor to $S$. This implies 
that the elements of $S$ (and hence of $S^m$) are of the form  $i\cdot d$ where $i\in{\mathbb Z}$. From Claim 1 it follows that the value of $l$ does not increase if we replace $S$ by $S^2$. Since $\max(S)$ certainly increases it follows that we can assume (by replacing $S$ by $S^m$ for sufficiently large $m$) that $\max(S)\geq 4l$. 
Further, by applying iteratively Claims 1 and 2 to $S'=S\cap[\min(S)+l^{-},\max(S)-l^{+}]_d$ we conclude that $(S')^m$ is an arithmetic $d$-progression whenever $m\geq 2^n$, namely $(S')^{m}=[m(\min(S)+l^{-}),m(\max(S)-l^{+})]_d$ . Since $S'\subseteq S$ it follows that $S^m$ contains $(S')^m$.  Now assume that $m$ is a multiple of $4$ and set $r=m\cdot\max(S)/4d$. It follows from $\max(S)\geq 4l$ that
$rd\geq ml^{-}$ and hence that $rd\in (S')^m$. It is shown in the same manner that $3rd\in {(S')}^m$. In summary, we have that $S^m$ and $r$ satify the hypothesis of 
Corollary~\ref{co:largeprogression}. Hence, every $d$-progression $\Diff_T$ where $|T| \leq r$ is pp-definable in $({\mathbb Z}; \suc, \Diff_S)$. The statement follows, because $r$ can be made arbitrarily large by increasing $m$. 
\end{Proof}


\begin{Lemma}\label{lem:not-progression-hard}
Suppose that $S$ is finite, but not an arithmetic $d$-progression, for any $d>0$. Then $\Csp(\Z;\suc, \mathrm{Diff}_S)$ is NP-hard. 
\end{Lemma}
\begin{Proof}
Let $S = \{a_1,\dots,a_k\}$ be such that $a_1 <\dots < a_k$. 
Let $d$ be the gcd of all $a_i-a_j$ with $i,j\in\{1,\dots,k\}$. By Lemma~\ref{lem:generatedprogression}, 
for all $i < k$ the relation 
$\mathrm{Diff}_{[a_{i-1},a_i]_d}$ is pp-definable in $\Gamma$. 
Then we obtain
$ \mathrm{Diff}_{\{a_{i-1},a_i\}}$ as $ \mathrm{Diff}_{[a_{i-1},a_i]_d}\cap  \mathrm{Diff}_S$. Since $S$ 
is not an arithmetic $d$-progression, 
for any $d > 0$, there exists an $i$ 
such that $a_i-a_{i-1} \neq a_{i+1}-a_i$. The result then follows from 
Lemma~\ref{lem:basichardcase}.
\end{Proof}

\begin{Lemma}\label{lem:two-progressions}
Let $S$ be an arithmetic $d$-progression
and $S'$ be an arithmetic $d'$-progression
such that $|S|>1$, $|S'|>1$,
and $d \neq d'$. Then
$\Csp(\Z;\suc, \mathrm{Diff}_S, \mathrm{Diff}_{S'})$ is NP-complete. 
\end{Lemma}
\begin{Proof}
We can pp-define $ \mathrm{Diff}_{\{a,a+d\}}$ and $ \mathrm{Diff}_{\{b,b+d'\}}$, normalize using $\suc$, and then apply Lemma~\ref{lem:basichardcase}.
\end{Proof}

As per the previous lemmas, we can now restrict 
our study to the situation where all binary relations
that are pp definable in $\Gamma$ are $d$-progressions, for some fixed $d$. We next treat
the case where $d=1$. 
The following technical lemma plays an important role in the sequel.

\begin{Lemma}
\label{lem:PPN-technical-1regime}
Suppose that $\Gamma$ contains
a non-trivial $1$-progression and 
a relation $R$ that is not $2$-decomposable, and that all binary relations pp-definable in $\Gamma$ are $1$-progressions. Then there exists $d \in \{-1,1\}$
such that $\Gamma$ pp-defines for each $m \geq 3$ a relation $T_m$ of arity $m'\geq m$ with
\[
\begin{array}{rll}
\leftarrow - \ m \ - \rightarrow & & \\
(d,0,\ldots,0,0 & ,0, \ldots,0) & \\
(0,d,\ldots,0,0 & ,0, \ldots,0) & \\
\vdots & \vdots & \\
(0,0,\ldots,d,0 & ,0, \ldots,0) &\\
(0,0,\ldots,0,d & ,0, \ldots,0) & \in T_m\\
\hline
(0,0,\ldots,0,0 & ,0, \ldots,0) & \notin T_m.\\
\end{array} 
\]
\end{Lemma}
\begin{Proof}
Assume $\Gamma$ has a relation $R$ that is not 2-decomposable. 
By replacing $R$ with a projection of $R$ to a subset of the arguments, we can assume
that $R$ has arity $r \geq 3$ and is not $(r-1)$-decomposable. This implies
that there exists a tuple
$(a_1,\dots,a_r) \notin R$ such that for all $i \in \{1,\dots,r\}$
there exists an integer $p_i$ such that $(a_1,\dots,a_{i-1},p_i,a_{i+1}, \dots, $ $a_r) \in R$. Replacing $R$ by the relation defined by the pp-formula
$$\exists y_1,\dots,y_r \big (\bigwedge_{i \in \{1,\dots,r\}} (y_i = x_i + a_i) \wedge R(y_1,\dots,y_r) \big )$$
we can further assume that $a_i = 0$ for all $i \in \{1,\dots,r\}$. 
Furthermore, we can also assume that $p_i \in \{-1,1\}$ 
for all $i \in \{1,\dots,r\}$. To see this, observe that $p_i \neq 0$ by assumption; if $p_i > 0$, choose $p_i$ minimal, if $p_i < 0$, choose $p_i$ maximal. 
Let $R_i$ be 
$\mathrm{Diff}_{[0,p_i-1|1]}$ if $p_i > 0$,
and let $R_i$ be 
$\mathrm{Diff}_{[p_i+1,0|1]}$ if $p_i < 0$. 
Note that by Lemma~\ref{lem:generatedprogression}, 
the relation $R_i$ has a pp
definition $\phi_i$ in $\Gamma$.
Now the formula
$$\theta:=\exists y_1,\dots,y_r \big(\bigwedge_{i \in \{1,\dots,r\}} \phi_i(x_i,y_i) \wedge R(y_1,\dots,y_r)\big) \; .$$
defines a relation where $p_i \in \{-1,1\}$ for
all $i \in \{1,\dots,r\}$. 

Let $P$ be the set of all $i \in \{1,\dots,r\}$ such that 
the tuple with a $1$ at the $i$-th position and $0$ everywhere else is in $R$. Likewise, let
$N$ be the set of all $i \in \{1,\dots,r\}$ such that 
the tuple with a $-1$ at the $i$-th position and $0$ everywhere else is in $R$.


{\bf Case 1.}
Suppose one of $P$ or $N$ is empty, \mbox{w.l.o.g.} $N$. 
Note that the relations 
$\Diff_{\{0,1\}}$ and 
$\Diff_{\{-1,0,1\}}$ are pp-definable in $\Gamma$ by Lemma~\ref{lem:generatedprogression}, so we may use them in pp-formulas over $\Gamma$. 
Define $\chi_r := \theta$, and
inductively define the pp-formula
$$\chi_j(x_1,\dots,x_{j-1}) := \exists x_j \, \big ( \chi_{j}(x_1,\dots,x_j) \wedge \Diff_{\{0,1\}}(x_r,x_{j+1}) \big)$$
for all $j \in \{3,\dots,r-1\}$.   
Define $N_j$ as the set of indices $i$ such that the tuple that contains 
$-1$ at the $i$th entry and $0$ otherwise
is in the relation defined by $\chi_j$.
We define $P_j$ analogously.
Note that $P_j$ is non-empty
for every $j \in \{3,\dots,r-1\}$.
If $N_j \neq \emptyset$, jump to
Case 2.  
 

So assume that $N_3$ is empty. Then the relation defined by 
\begin{align*}
\exists x_r,y_r,z \, \big ( & \theta(x_1,\dots,x_r) \wedge
\theta(y_1,\dots,y_r) \wedge \chi_3(x_r,y_r,z) \\
\wedge & \Dist_{\{0,1\}}(x_1,x_r) \wedge \Dist_{\{0,1\}}(y_1,y_r) 
\wedge \Dist_{\{0,1\}}(x_1,z)
\wedge \Dist_{\{0,1\}}(x_r,z) \big )
\end{align*}
has the required properties for 
$T_{2(r-1)}$: if all of $x_1,\dots,x_{r-1},y_1,\dots,y_{r-1}$ are equal to zero, then $x_r = y_r = 1$ because
of the conjuncts $\theta(x_1,\dots,x_r)$
and $\theta(y_1,\dots,y_r)$, the assumption that $N$ is empty, and the conjuncts $\Dist_{\{0,1\}}(x_1,x_r)$ 
and $\Dist_{\{0,1\}}(y_1,y_r)$.  
Hence, 
$z = 2$ 
because of the conjuncts 
$\chi_3(x_r,y_r,z)$ and $\Dist_{\{0,1\}}(x_r,z)$, 
in contradiction to $\Dist_{\{0,1\}}(x_1,z)$. 
On the other hand, if $x_i$ is set to $1$ for $1 \leq i < r$ and all other variables
in $\{x_1,\dots,x_{r-1},y_1,\dots,y_{r-1}\}$ are set to $0$, 
then $y_r$ can be set to $1$, $x_r$ can be set to $0$,
and $z$ can be set to $0$, and this satisfies
all conjuncts of the formula.  

Iterating this construction
we obtain pp-definitions of relations
$T_m$, for arbitrary $m >r$, 
with the required properties. 

\vspace{.2cm}
{\bf Case 2.} Suppose that both $P$ and $N$ are non-empty and let $i \in P$ and $j \in N$. 
Consider the pp-formula $\phi(x_1,\dots,x_{i-1},x_{i+1},\dots,x_r,y_1,\dots,y_{j-1},y_{j+1},\dots,y_r)$ given by
\begin{align*}
\exists x,y \, \big( \suc(y,x) \wedge  
R(x_1,\dots,x_{i-1},& x,x_{i+1},\dots,x_r) \\
\wedge \; R(y_1,\dots,y_{j-1},& y,y_{j+1},\dots,y_r) \\
 \wedge \, \Diff_{\{-1,0,1\}}(x,x_1) \, & \wedge \, \Diff_{\{-1,0,1\}}(y,y_1) \big) \; .
\end{align*}
Assume \mbox{w.l.o.g.} that there is $k \in P \setminus \{i\}$, \mbox{i.e.}, $|P|>1$.
Reordering the arguments of the relation 
defined by $\phi$ such that the variables
 $y_i,y_k$, and $x_k$ correspond to the 
 first $m = 3$ arguments, we obtain 
  a relation $T_3$ of arity $m' := r \geq 3$ with
the desired properties. To see this, consider the case that all variables from 
$V := \{x_1,\dots,x_{i-1},x_{i+1},\dots,x_r,y_1,\dots,y_{j-1},y_{j+1},\dots,j_r\}$ are set to $0$. 
Then the first conjunct of $\phi$ implies that $x \neq 0$ and the second conjunct that $y \neq 0$,
and then the conjuncts $\Diff_{\{-1,0,1\}}(x,x_k)$ and $\Diff_{\{-1,0,1\}}(y,y_k)$ 
are inconsistent with $\suc(y,x)$.
On the other hand, 
\begin{itemize}
\item if $y_i = 1$ and all other variables in $V$ are set to $0$, then setting $x$ to $1$ and $y$ to $0$ satisfies all conjuncts of $\phi$. 
\item if $y_k = 1$ and all other variables in $V$ are set to $0$, then setting $x$ to $1$ and $y$ to $0$ satisfies all conjuncts of $\phi$. 
\item if $x_k = 1$ and all other variables in $V$ are set to $0$, then setting $x$ to $0$ and $y$ to $-1$ satisfies all conjuncts of $\phi$. 
\end{itemize}
Informally, this can be illustrated by the following table. 
\[
\begin{array}{cc|ccc}
y & x & y_i & y_k & x_k \\
\hline
 0 & 1 & 1 & 0 & 0 \\
 0 & 1 & 0 & 1 & 0 \\
-1 & 0 & 0 & 0 & 1 \\
\hline
  &  & 0 & 0 & 0 \\
\end{array}
\]

For the case $m>3$, we can iterate this construction, replacing $R$ by the relation $R'$ defined by $\phi$. To see this, note that
the set $P$ redefined with respect to $R'$
contains the entries for the variables
$y_i, y_k, x_k$, and thus $|P|>2$. Moreover, because all conjuncts of $\phi$ are true under the assignment $x_j=-1$, $y=-1$, $x=0$, and all other variables set to $0$, the set $N$ redefined with respect to $R'$ contains the entry
for the variable $x_j$, so that $|N|\geq 1$. 
\end{Proof}


\vspace{.2cm}
The following proposition replaces
the proof of Theorem~31 in the conference version of this paper 
which contained an important error.

\begin{Proposition}
\label{prop:new-with-1-progression}
Suppose that $\Gamma$ contains the relation $\suc$ and a non-trivial $1$-progression. 
Then $\Gamma$ is preserved by one of max or min;
or $\Csp(\Gamma)$ is NP-hard. 
\end{Proposition}
\begin{Proof}
If $\Csp(\Gamma)$ is not NP-hard, 
then by Lemmas~\ref{lem:two-progressions} and~\ref{lem:not-progression-hard}
we can assume that all the
binary relations with a pp-definition in $\Gamma$ are $1$-progressions. 
It follows that every 
$2$-decomposable relation 
is preserved by both max and min. If every relation
is preserved by both max and min, then we are done,
so assume in the following 
that $\Gamma$ has a relation 
$R$ that is not 2-decomposable. We now find ourselves with the preconditions of Lemma~\ref{lem:PPN-technical-1regime}. \mbox{W.l.o.g.} assume $d=1$ (for $d=-1$ we potentially generate min instead of max in the following).

We now claim that 
for every finite set $[-n,n] := \{-n,-n+1,\dots,n-1,n\} \subset\Z$, the operation $\max\colon [-n,n]^2\to [-n,n]$ is a polymorphism of the substructure of $\Gamma$ induced by $[-n,n]$, which we denote
by $\Gamma[-n,n]$ in the following. 
If $\Csp(\Gamma[-n,n], 0)$ were NP-hard, then $\Csp(\Gamma)$ would also be NP-hard. Indeed, we have by Lemma~\ref{lem:generatedprogression}
that the $1$-progression $\Diff_S$ with $S=\{-1,0,1\}$ is pp-definable in $\Gamma$. As a consequence, the progression $\Diff_T$ with $T=[-n,n]$
is pp-definable in $\Gamma$ by a pp formula of size $O(n)$. Our reduction from $\Csp(\Gamma[-n,n], 0)$ to $\Csp(\Gamma)$
works as follows: from an input $\Phi$ of $\Csp(\Gamma[-n,n], 0)$ with variable set $V$, create the instance $\Psi := \exists z \big( \Phi\land \bigwedge_{v\in V} \Diff_T(z,v) \big )$
of $\Csp(\Gamma)$ where each atom $v=0$ in $\Phi$ is replaced by $v=z$. Note that $\Psi$ can be computed in polynomial time from $\Phi$.
By transitivity of the structure $\Gamma$, we have that $\Phi$ is true in $(\Gamma[-n,n], 0)$ iff $\Psi$ is true in $\Gamma$, thus proving our claim that $\Csp(\Gamma)$ is NP-hard.

Thus we may assume that each $\Csp(\Gamma[-n,n], 0)$ is not NP-hard. 
Note that all polymorphisms $f$ of $(\Gamma[-n,n],0)$ are
\emph{idempotent}, i.e., 
$f(x,\dots,x) = x$ for all $x \in [-n,n]$,  
since $0$ and $\suc$ are in the language of $\Gamma$. 
It is known from the theory of finite-domain constraint satisfaction, by 
a combination of a result of 
Jeavons, Bulatov, and Krokhin~\cite{JBK} (Corollary 7.3) and of
Mar\'oti and McKenzie~\cite{MarotiMcKenzie} (Theorem 1.1; in order to match the terminology between these papers, we refer to Section 3.2 in the survey article~\cite{Bulatov-Valeriote}), that
in this case $\Gamma[-n,n]$ has an (idempotent) \emph{weak near-unanimity} polymorphism $f_n$,  
that is, $f_n$ has arity $k \geq 2$ and satisfies for 
all $x,y \in [-n,n]$ the equation
$$f_n(y,x,\dots,x) = f_n(x,y,x,\dots,x) = \dots = f_n(x,\dots,x,y) \; .$$

Fix now an integer $n$, and let $k$ be the arity of $f_n$. We will prove that $f_n(x,\dots,x,y) = \max(x,y)$ for all $x,y \in [-n,n]$. 
Since $f_n(0,\dots,0)=0$, and $f_n$ must preserve $\mathrm{Diff}_{[0,1|1]}$, we deduce that $f_n(a_1,\dots,a_k) \in \{0,1\}$
for all $a_1,\dots,a_k \in \{0,1\}$. 
Consider now the following $k$ tuples contained in $T_k$:
$$\begin{matrix}
	(1, & 0, & 0, & \dots & 0, & 0, & \dots & 0),\\
	(0, & 1, & 0, & \dots & 0, & 0, & \dots & 0),\\
	& & \ddots & & &\vdots & &\\
	(0, & 0, & 0, & \dots & 1, & 0, & \dots & 0).
\end{matrix}
$$
Since $f_n$ is an idempotent weak near-unanimity, by applying $f_n$ to these tuples we obtain $(a,\dots,a,0,\dots,0)$,
where $a=f_n(0,\dots,0,1)$
Moreover, $a \neq 0$ since $(0,\dots,0) \notin T_k$. We obtain that $f_n(a_1,\dots,a_k)=1$ whenever
exactly one of the $a_i$ equals $1$ and the other $a_i$ equal
$0$. By preservation of $\suc$, we also obtain that 
$f_n(2,1,\dots,1)=2$. Since $(1,2),(1,1),\dots,(1,1),(0,1) \in \Diff[0,1|1]$ we have $f_n(1,\dots,1,0) \in \{1,2\}$. 
We have already observed that $f_n(1,\dots,1,0) \in \{0,1\}$, and hence $f_n(1,\dots,1,0)=1$. 
Analogously, the value of $f_n$ is determined when
all arguments are $1$, except for one that equals $0$. 
It follows by preservation of $\suc$ 
that for all $p,q \in [-n,n]$ 
with $|p-q|=1$ 
$$f_n(p,\dots,p,q)=\max(p,q) \, .$$
We now aim to prove that this holds for all $p,q \in [-n,n]$. 
Assume by induction on $t$ that we have 
the result for $|p-q| \leq t$, and that we want to show it for $|p-q| = t+1$. 
By inductive hypothesis we have $f(0,\dots,0,t) = t$, and $f(0,\dots,0,t+1) \in 
\{t,t+1\}$ by preservation of $\mathrm{Diff}_{[0,1|1]}$. 
 We now apply $f_n$ componentwise to the 
following tuples from $T_k$: 
 \begin{align*}
 (0,1,0,\dots,0,0) & \in T_k \\
 & \vdots \\
 (0,1,0,\dots,0,0) & \in T_k \\
 (t+1,t,\dots,t,t) & \in T_k 
 \end{align*}
Since $f(0,\dots,0,t+1) \in \{t,t+1\}$
and $(t,\dots,t) \notin T_k$, 
we obtain $f(0,\dots,0,t+1) = t+1$. 
A similar argument shows that $f_n(t+1,\dots,t+1,0) = t+1$. 
This implies the inductive claim for $p,q \in [-n,n]$ with 
$|p-q| = t+1$ because $f_n$ is idempotent and preserves $\suc$. 
Since $\max$ agrees on each finite set with a polymorphism of
$\Gamma[-n,n]$, it follows that $\max$ is a polymorphism of $\Gamma$. 
\end{Proof}

\vspace{0.2cm}
{\bf Proof of Theorem~\ref{thm:class}.}
If $\Gamma$ is preserved by a modular max or modular min, then $\Csp(\Gamma)$ is in P by Theorem~\ref{thm:tractability-modular-semilattice}. 
So suppose that this is not the case.
By Lemma~\ref{lem:positive-not-modular-binary-relation} there is a binary relation $R$ 
with a pp-definition in $\Gamma$ but not in $(\Z;\suc)$.
If $R$ is not a $d$-progression for any $d\geq 1$,
then $\Csp(\Gamma)$ is NP-hard by Lemma~\ref{lem:not-progression-hard}.
So suppose that $R$ is a $d$-progression.
If there is a non-trivial $d'$-progression
with $d' \neq d$ then $\Csp(\Gamma)$ is NP-hard by Lemma~\ref{lem:two-progressions}. 
So suppose that all non-trivial
binary relations with a pp-definition
in $\Gamma$ are $d$-progressions. 
Then the conditions of Lemma~\ref{lem:reduction-all-equal-mod-d}
apply and we can assume without loss
of generality that $\Gamma$
is $d$-nice.

The relation defined by $y=s^d(x)$ is pp-definable in $\Gamma$, 
and by adding this relation we see that 
$\Gamma/d$ contains $\suc$ and so is as in Proposition~\ref{prop:new-with-1-progression}. 
By Lemma~\ref{lem:modular-max-max},
the $d$-nice structure $\Gamma/d$ is not preserved by $\max$ or $\min$,  and hence Proposition~\ref{prop:new-with-1-progression} implies that $\Csp(\Gamma/d)$ is NP-hard. Now we reduce $\Csp(\Gamma/d)$ to $\Csp(\Gamma)$ to prove the latter is also NP-hard. Recall that $D$ denotes the largest distance
 in the Gaifman graph of $\Gamma$ (Notation~\ref{notation:distances}). Note that an instance of $\Gamma$ on $n$ variables has a solution if and only if it has a solution in the interval $[0,Dn]$. From an instance $\Phi$ of $\Csp(\Gamma/d)$ we build an instance $\Psi$ of $\Csp(\Gamma)$. To build $\Psi$ from $\Phi$, we augment with a new variable $z$ as well as $Dn$ new variables $x_1\ldots,x_{Dn}$ for each extant variable $x$ of $\Psi$. Then $\Psi$ is as $\Phi$ but with the additional constraints $\Dist_{[0,Ddn]_d}(x,z)$, where we define $\Dist_{[0,Ddn]_d}(x,z)$ by $\Dist_{[0,d]_d}(x,x_1) \wedge \Dist_{[0,d]_d}(x_1,x_2) \wedge \ldots \wedge \Dist_{[0,d]_d}(x_{Dn},z)$. It is straightforward to see that $\Gamma/d \models \Phi$ if and only if $\Gamma \models \Psi$ and the result follows.  \qed 


\vspace{1cm}

{\bf Proof of Theorem~\ref{thm:main}.}
Suppose that $\Gamma$ does not have a finite core. 
Let $\Delta$ be the structure as described in Theorem~\ref{thm:def-succ};
that is, $\Delta$ is a connected finite-degree structure with a first-order definition in $(\Z;\suc)$ such that 
there is a homomorphism $e$ from $\Gamma$ to 
$\Delta$ and a homomorphism $i$ from $\Delta$ to $\Gamma$. Clearly, $\Csp(\Gamma)$ and $\Csp(\Delta)$ are the same problem. Theorem~\ref{thm:def-succ} asserts that the relation 
$\suc$ is pp-definable in $\Delta$ unless $\Csp(\Delta)$ 
(and $\Csp(\Gamma)$) is NP-hard. 
If $\suc$ is pp-definable in $\Delta$, then
the CSP of the expansion of $\Delta$ by the successor relation
has the same complexity as $\Csp(\Delta)$. 
Theorem~\ref{thm:class} implies that $\Delta$ has
a modular max or modular min, and $\Csp(\Delta)$ and $\Csp(\Gamma)$ are in P, or $\Csp(\Delta)$ and $\Csp(\Gamma)$ are NP-hard. 
\qed

\section{Concluding Remarks}
\label{sect:concl}
Structures $\Gamma$ with a first-order definition in $({\mathbb Z}; \suc)$
have a \emph{transitive} automorphism group, i.e., for every $x,y \in {\mathbb Z}$ there is an automorphism of $\Gamma$ that maps $x$ to $y$. We call such structures $\Gamma$ \emph{transitive} as well.
It is well-known and easy to prove (see e.g.~\cite{HNBook}) that a finite core of a transitive structure is again transitive.
Our main result thus implies that a complete complexity classification for distance CSPs 
follows from a complexity classification for CSPs whose template is a finite transitive core.
In general, the complexity of CSPs for
finite transitive templates has not yet been classified. The
following is known.

\begin{Theorem}[of~\cite{JBK,BartoKozikLICS10}]
Let $\Delta$ be finite.
If $\Delta$ has no polymorphism $f$ of arity $n \geq 2$ satisfying
$$\forall x_1,\dots,x_n. f(x_1,\dots,x_n) = f(x_2,\dots,x_n,x_1)$$
then $\Csp(\Delta)$ is NP-complete.
\end{Theorem}

The following conjecture is widely believed in the area. 

\begin{Conjecture}[of~\cite{JBK,BartoKozikLICS10}]
Let $\Delta$ be finite. If $\Delta$
has for some $n \geq 2$ an $n$-ary polymorphism $f$ satisfying
$$\forall x_1,\dots,x_n. f(x_1,\dots,x_n) = f(x_2,\dots,x_n,x_1)$$
then $\Csp(\Delta)$ is in P. 
\end{Conjecture}

The authors believe that this conjecture might be easier to show for \emph{transitive}
finite structures $\Gamma$. 

We mention that recently, the (infinite) lattice of structures over $\mathbb Z$
with a first-order definition in $(\Z;\suc)$ considered up
first-order interdefinability has been described~\cite{SemenovSoprunovUspensky}. 



The general classification program of distance CSPs, in which one relaxes the requirement of local finiteness, but still insist on a finite signature, has recently been completed~\cite{BodMarMot}.
All distance CSPs are in P or NP-complete,
or they are homomorphically equivalent to 
transitive finite structures, in which case a general 
complexity classification is not known. 
The quest for understanding the infinite signature case, under encodings in disjunctive normal form, is ongoing. 

\section*{Acknowledgements}
We are grateful to several anonymous reviewers
for their valuable comments.  

\bibliographystyle{plain}
\bibliography{local}

\def\cprime{$'$}
\begin{thebibliography}{10}

\bibitem{BartoKozikLICS10}
Libor Barto and Marcin Kozik.
\newblock New conditions for {T}aylor varieties and {CSP}.
\newblock In {\em Proceedings of LICS}, pages 100--109, 2010.

\bibitem{Cores-journal}
Manuel Bodirsky.
\newblock Cores of countably categorical structures.
\newblock {\em Logical Methods in Computer Science}, 3(1):1--16, 2007.

\bibitem{BodChenPinsker}
Manuel Bodirsky, Hubie Chen, and Michael Pinsker.
\newblock The reducts of equality up to primitive positive interdefinability.
\newblock {\em Journal of Symbolic Logic}, 75(4):1249--1292, 2010.

\bibitem{BodDalMarPin}
Manuel Bodirsky, V\'{\i}ctor Dalmau, Barnaby Martin, and Michael Pinsker.
\newblock Distance constraint satisfaction problems.
\newblock In Petr Hlinen\'y and Anton\'in Kucera, editors, {\em Proceedings of
  Mathematical Foundations of Computer Science}, Lecture Notes in Computer
  Science, pages 162--173. Springer Verlag, August 2010.

\bibitem{BodirskyGrohe}
Manuel Bodirsky and Martin Grohe.
\newblock Non-dichotomies in constraint satisfaction complexity.
\newblock In Luca Aceto, Ivan Damgard, Leslie~Ann Goldberg, Magn\'us~M.
  Halld\'orsson, Anna Ing\'olfsd\'ottir, and Igor Walukiewicz, editors, {\em
  Proceedings of the International Colloquium on Automata, Languages and
  Programming (ICALP)}, Lecture Notes in Computer Science, pages 184 --196.
  Springer Verlag, July 2008.

\bibitem{BodHilsMartin}
Manuel Bodirsky, Martin Hils, and Barnaby Martin.
\newblock On the scope of the universal-algebraic approach to constraint
  satisfaction.
\newblock In {\em Proceedings of the Annual Symposium on Logic in Computer
  Science (LICS)}, pages 90--99. IEEE Computer Society, July 2010.

\bibitem{tcsps-journal}
Manuel Bodirsky and Jan K\'ara.
\newblock The complexity of temporal constraint satisfaction problems.
\newblock {\em Journal of the ACM}, 57(2):1--41, 2009.
\newblock An extended abstract appeared in the Proceedings of the Symposium on
  Theory of Computing (STOC'08).

\bibitem{BodMacpheTha}
Manuel Bodirsky, Dugald Macpherson, and Johan Thapper.
\newblock Constraint satisfaction tractability from semi-lattice operations on
  infinite sets.
\newblock {\em Transaction of Computational Logic (ACM-TOCL)}, 14(4):1--30,
  2013.

\bibitem{BodMarMot}
Manuel Bodirsky, Antoine Mottet, and Barnaby Martin.
\newblock Constraint satisfaction problems over the integers with successor.
\newblock In {\em Proceedings of ICALP}, 2015.
\newblock ArXiv:1503.08572.

\bibitem{BodirskyNesetrilJLC}
Manuel Bodirsky and Jaroslav Ne\v{s}et\v{r}il.
\newblock Constraint satisfaction with countable homogeneous templates.
\newblock {\em Journal of Logic and Computation}, 16(3):359--373, 2006.

\bibitem{BodPin-Schaefer-both}
Manuel Bodirsky and Michael Pinsker.
\newblock Schaefer's theorem for graphs.
\newblock {\em Journal of the ACM}, 62(3):\#19, 52 pages, 2015.
\newblock A conference version appeared in the Proceedings of STOC 2011, pages
  655--664.

\bibitem{Bulatov-Valeriote}
A.~Bulatov and M.~Valeriote.
\newblock Results on the algebraic approach to the csp.
\newblock {\em Complexity of Constraints: An Overview of Current Research
  Themes}, pages 68--92, 2008.
\newblock Springer Verlag.

\bibitem{JBK}
Andrei~A. Bulatov, Andrei~A. Krokhin, and Peter~G. Jeavons.
\newblock Classifying the complexity of constraints using finite algebras.
\newblock {\em SIAM Journal on Computing}, 34:720--742, 2005.

\bibitem{Oligo}
Peter~J. Cameron.
\newblock {\em Oligomorphic permutation groups}.
\newblock Cambridge University Press, Cambridge, 1990.

\bibitem{CSPSurveys}
Nadia Creignou, Phokion~G. Kolaitis, and Heribert Vollmer, editors.
\newblock {\em Complexity of Constraints - An Overview of Current Research
  Themes [Result of a Dagstuhl Seminar]}, volume 5250 of {\em Lecture Notes in
  Computer Science}. Springer, 2008.

\bibitem{diestel}
Reinhard Diestel.
\newblock {\em Graph Theory}.
\newblock Springer--Verlag, New York, 2005.
\newblock 3rd edition.

\bibitem{FederVardi}
Tom\'as Feder and Moshe~Y. Vardi.
\newblock The computational structure of monotone monadic {SNP} and constraint
  satisfaction: {a} study through {D}atalog and group theory.
\newblock {\em {SIAM} Journal on Computing}, 28:57--104, 1999.

\bibitem{ModelTheoryShawn}
Shawn Hedman.
\newblock {\em A First Course in Logic: An Introduction to Model Theory, Proof
  Theory, Computability, and Complexity (Oxford Texts in Logic)}.
\newblock Oxford University Press, Inc., New York, NY, USA, 2004.

\bibitem{HellNesetril}
Pavol Hell and Jaroslav Ne\v{s}et\v{r}il.
\newblock On the complexity of {H}-coloring.
\newblock {\em Journal of Combinatorial Theory, Series B}, 48:92--110, 1990.

\bibitem{HNBook}
Pavol Hell and Jaroslav Ne\v{s}et\v{r}il.
\newblock {\em Graphs and Homomorphisms}.
\newblock Oxford University Press, Oxford, 2004.

\bibitem{Hodges}
Wilfrid Hodges.
\newblock {\em A shorter model theory}.
\newblock Cambridge University Press, Cambridge, 1997.

\bibitem{JeavonsClosure}
Peter Jeavons, David Cohen, and Marc Gyssens.
\newblock Closure properties of constraints.
\newblock {\em Journal of the ACM}, 44(4):527--548, 1997.

\bibitem{Marker}
David Marker.
\newblock {\em Model Theory: An Introduction}.
\newblock Springer, New York, 2002.

\bibitem{MarotiMcKenzie}
M.~Mar\'oti and R.~McKenzie.
\newblock Existence theorems for weakly symmetric operations.
\newblock {\em Algebra Universalis}, 59(3), 2008.

\bibitem{SemenovSoprunovUspensky}
Alexei Semenov, Sergey Soprunov, and Vladimir Uspensky.
\newblock The lattice of definability. {O}rigins, recent developments, and
  further directions.
\newblock In {\em Computer Science Russia}, volume 8476 of {\em Lecture Notes
  in Computer Science}, pages 23--38, 2014.

\end{thebibliography}
\end{document}